\documentclass[%
 reprint,
 amsmath,amssymb,
 aps,
prb,
]{revtex4-2}

\usepackage{graphicx}% Include figure files
\usepackage{dcolumn}% Align table columns on decimal point
\usepackage{bm}% bold math
\usepackage[T1]{fontenc}

\usepackage[version=3]{mhchem, xcolor} % Formula subscripts using \ce{}
\usepackage{siunitx}
\usepackage{physics}
\AtBeginDocument{\RenewCommandCopy\qty\SI}
\DeclareSIUnit\angstrom{\text {Å}}

\usepackage{xspace}
\newcommand{\tadah}{\textit{Ta-dah!}\xspace}
\newcommand{\angstrom}{\AA\, }

\newcommand{\ALPHA}{$\alpha$-\ce{N2}\xspace}
\newcommand{\BETA}{$\beta$-\ce{N2}\xspace}
\newcommand{\GAMMA}{$\gamma$-\ce{N2}\xspace}
\newcommand{\DELTA}{$\delta$-\ce{N2}\xspace}
\newcommand{\DELTALOC}{$\delta^*$-\ce{N2}\xspace}
\newcommand{\EPSILON}{$\epsilon$-\ce{N2}\xspace}
\newcommand{\IOTA}{$\iota$-\ce{N2}\xspace}
\newcommand{\LAMBDA}{$\lambda$-\ce{N2}\xspace}

\begin{document}

\preprint{APS/123-QED}

\title{Understanding solid nitrogen through machine learning simulation}

\author{Marcin Kirsz}
\affiliation{Centre for Science at Extreme Conditions and School of Physics and Astronomy, University of Edinburgh, Edinburgh, U.K.}

\author{Ciprian G. Pruteanu}
\affiliation{Centre for Science at Extreme Conditions and School of Physics and Astronomy, University of Edinburgh, Edinburgh, U.K.}

\author{Peter I. C. Cooke}
\affiliation{Department of Materials Science \& Metallurgy, University of Cambridge, Cambridge, U.K.}

\author{Graeme J. Ackland}\email{gjackland@ed.ac.uk}
\affiliation{Centre for Science at Extreme Conditions and School of Physics and Astronomy, University of Edinburgh, Edinburgh, U.K.}

\date{\today}

\begin{abstract}
We construct a fast, transferable, general purpose,  machine-learning interatomic potential suitable for large-scale simulations of \ce{N2}. The potential is trained only on high quality quantum chemical molecule-molecule interactions, no condensed phase information is used. The potential reproduces the experimental phase diagram including the melt curve and the molecular solid phases of nitrogen up to \qty{10}{GPa}.
This demonstrates that many-molecule interactions are unnecessary to explain the condensed phases of \ce{N2}.
With increased pressure, transitions are observed from cubic (\ALPHA), which optimises quadrupole-quadrupole interactions, through tetragonal (\GAMMA) which allows more efficient packing, through to monoclinic (\LAMBDA) which packs still more efficiently. On heating, we obtain the hcp 3D rotor phase (\BETA) and, at pressure, the cubic \DELTA phase which contains both 3D and 2D rotors, tetragonal \DELTALOC phase with 2D rotors and the rhombohedral \EPSILON. Molecular dynamics demonstrates where these phases are indeed rotors, rather than frustrated order. The model does not support the existence of the wide range of bondlengths reported for the complex \IOTA phase. The thermodynamic transitions involve both shifts of molecular centres and rotations of molecules. We simulate these phase transitions between finding that the onset of rotation is rapid whereas motion of molecular centres is inhibited and the cause of the observed sluggishness of transitions.  Routine density functional theory calculations give a similar picture to the potential.

%\begin{description}
%You may use the \texttt{description} environment to %structure your abstract;
%use the optional argument of the \verb+\item+ command to %give the category of each item. 
%\end{description}
\end{abstract}

\maketitle

%\tableofcontents

\section{Introduction}
Bonding in solid nitrogen is extremely inhomogeneous.
On one hand, the triple bond of nitrogen is very strong with a dissociation energy of \qty{9.72}{eV/molecule} \cite{Strak2007}. On the other, interactions between molecules are weak, and the free energy differences between competing crystal structures can be extremely small, in the meV/molecule range.

The many ways to orient the \ce{N2} molecule mean that the pressure-temperature phase diagram of condensed nitrogen is extremely complex. The molecular stability of nitrogen persists to over \qty{100}{GPa}, so crystal phases below that pressure involve ordering of well-defined molecules.
Above \qty{100}{GPa}, molecular bond dissociates and either an amorphous solid \cite{Goncharov2000, Eremets2001} or the crystalline cubic gauche structure \cite{Eremets2004} is observed to form depending on exact P, T conditions.

Even within the molecular limit, there is a surprisingly large diversity of crystal structures.  In addition to the melt, a series of six crystalline phases below \qty{10}{GPa} had been experimentally reported by multiple groups up to 2016 ($\alpha, \beta, \gamma, \delta, \delta^*$ and $\epsilon$ \cite{Scurlock1992, Eucken1916, Swenson1955, Stewart1956, Cromer1981, Schiferl1985, mills1986structures, Olijnyk1990, stinton2009crystal}), and more recently two more ($\lambda$ and $\iota$ \cite{frost2016novel, turnbull2018unusually}).  Transformations between them can be sluggish, so experiments tend to show considerable hysteresis.  With the exception of iota, the reported \ce{N2} bondlengths in the crystal phases are all close to \qty{1.09}{\angstrom}.

The accurate description of the potential energy surface of the system of interacting \ce{N2} molecules plays an essential role in governing its dynamics and properties. Solid nitrogen has been studied extensively with a variety of theoretical methods such as free electron gas model \cite{LeSarGordon1983N2DFT, LeSar1984ImprovedGasModelN2} (FEG), density functional theory (DFT) \cite{pickard2009high}, Møller–Plesset perturbation theory \cite{Erba2011} (MP), coupled cluster (CC) \cite{Hellmann2013}, Monte Carlo \cite{Belak1988, Belak1990, Mulder1996, Mulder1997} (MC), molecular dynamics (MD) and others \cite{Zunger1975, Murthy1980, NoseKlein1986, Etters1986}.

For the purpose of atomistic simulation a number of interatomic potentials (IP) have been developed. It was quickly recognised that the simple atom-atom pair potentials are insufficient \cite{Murthy1980}.
The solid phases of nitrogen have been the subject of extensive MD studies and the ability to reproduce various \ce{N2} phases, has been recognised in early reviews \cite{Zunger1975,Murthy1980}.
Interestingly, Nose and Klein showed that the $\alpha$ phase predicted using LJ atom-atom pairwise potentials was different from that when quadrupole interactions were added, and different again from a formal charge model.  \cite{NoseKlein1986}. 
They argued that the quadrupole-quadrupole interaction is unimportant for \ce{N2} at moderate pressure based on the similarity of the \DELTA phase to the ambient pressure crystal structure of oxygen. However their simple potential grossly overestimated the volume of the trigonal cell as compared with the experiment and is inadequate for other crystal phases.

While great progress has been made in identifying relevant physical mechanisms governing the behaviour of the nitrogen solid phases, an IP for nitrogen which explains the seven phases below \qty{10}{GPa} has remained elusive.
For a long time it seemed that only potentials specifically fitted to reproduce $\alpha\rightarrow\gamma$ could obtain it \cite{RaichMills,Mandell1974} and purpose-built potentials were needed to study different crystal phases \cite{KOBASHI198574, NoseKlein1986, Powell1989, Westerhoff1997}.

Several of the proposed \ce{N2} phases are believed to involve disordered or freely rotating molecules.  These two situations are difficult to resolve experimentally, but they constitute an ideal problem to tackle with MD.
However accurate IPs have remained the problem.
It is perhaps surprising that for such a fundamental system as \ce{N2} there is no general purpose transferable IP capable of reproducing well established experimental phases as well as the liquid state.
Currently, the NIST Interatomic Potentials repository lists just one 6-12 LJ potential with the following note: ``(...) its ability to model structures other than dimers is unknown'' \cite{Becker2013, Hale2018}.
A brief test verifies that it is rather insufficient for condensed phases.

Machine learning interatomic potentials (MLIP) trained on density functional theory data have become a go-to method for describing complex phases in condensed matter. 
We have developed a flexible machine learning package, \tadah, which enables us to implement this, alongside community codes CASTEP and LAMMPS \cite{Segall2002, Thompson2022LAMMPSScales}. Unfortunately, much understanding of the underlying chemistry is lost in the gigabytes of data needed for an accurate DFT calculation.  This is then compounded by deriving a potential via ``machine learning'', a process which is good at replicating nature, but not equivalent to ``researcher understanding''.  

Moreover, machine-learning models are often only reliable for interpolating within the regime where they are trained - successful extrapolation requires physical insight.

The Frenkel line for nitrogen was previously studied with a \tadah potential trained directly on CCSDT(Q) data \cite{Pruteanu2021}. Therein we introduced a self-teaching method for machine learning, in which a series of trial MLIPS are built, each of which is used to generate a training dataset for its successor. While the model was successful it is still just a black box with limited transferability beyond the training dataset.

Consequently, here we build a readily understandable model for nitrogen with greatly improved transferability.
We demonstrate that the low pressure \ce{N2} phases can be readily understood by two-molecule pairwise interactions, trained with no reference to condensed phase electronic structure calculation. 

This paper is structured as follows: first, we outline  the machine-learning procedure as implemented in the \tadah package; second, the methods are then developed to generate a model based on CCSDT(Q) calculations; third, the physical accuracy of the potential is assessed in large-scale simulations by comparison to the known experimental phase diagram of \ce{N2}.

\section{Potential Development}
The potential developed in this report makes use of our \tadah software which is publicly available at
\url{https://git.ecdf.ed.ac.uk/tadah} along with an extensive documentation.
The package is designed to assist in the development of custom-made MLIPs and deployment of those in LAMMPS\cite{Thompson2022LAMMPSScales} via a provided plugin.
\tadah is written in modern C++ and its modular structure allows rapid implementation and testing of new ideas, followed by seamless deployment to large-scale MD simulations.
The code provides an easy-to-use command line interface as well as C++ application programming interface for more advanced use.

The small energy differences between different competing solid phases require extremely accurate training datasets, including accurate dispersion forces as well as an adequate parametrisation procedure. \tadah incorporates a two-stage fitting procedure where the nonlinear hyperparameters in the model's descriptors are simultaneously optimised along with the usual machine-learning of model parameters with linear algebra. The detailed description of \tadah machine-learning and hyperparametrisation procedure is published elsewhere\cite{thesis}. Here we limit our discussion to physically meaningful detail.

The local energy of molecule $i$ is obtained by iterating over all of its nearest neighbouring molecules within a center of mass cutoff distance $r_c=\qty{12}{\angstrom}$ and summing over each molecule-molecule interaction in a pairwise fashion.
The total energy of the system, $E_{total}$ is then obtained by accumulating all local molecular energies $E_i$
\begin{equation}
    E_{total}=\sum_i E_i
\end{equation}

The atomic forces are readily available from the derivative of the total energy with respect to the atomic positions.
However, the force between bonded atoms is removed using the SHAKE algorithm \cite{RYCKAERT1977327} as implemented in LAMMPS such that the bond length is kept fixed at \qty{1.1014}{\angstrom}.  We note that the self-consistency loop in SHAKE is unnecessary for diatomic molecules, so the algorithm is much faster than for more complex molecules.

The choice of rigid bonds means the vibrational degree of freedom is not excited.  At \qty{2739}{cm^{-1}}, the mode is only excited at around \qty{4000}{K}, well above any temperatures considered here. 
Rigid bonding means that the potential has no spurious contribution to the heat capacity, as a flexible bond would. Nevertheless, the inclusion of the bond length in the descriptor implies that the bond energy varies with environment, so the functional form can fit the weakening of the triple bond with pressure.

\subsection{Descriptor and regression choice}

For the \ce{N2} molecular system each atom is permanently associated with a molecule.
The local atomic environment of each atom is captured by a combination of two-body (eq. \ref{eq::N2::blip_2b}) and many-body descriptors (eq. \ref{eq::N2::MB}) as implemented in \tadah.
To represent the interaction between molecules $i$ and $j$ we choose descriptors to have chemically intuitive meaning.

The two-body terms can be loosely associated with short-ranged repulsion and van der Waals interactions while the many-body term captures complex electric multipole interactions. In practice both terms are fitted simultaneously by polynomial regression.

This means that the energy and forces for each molecule can be written as a sum over interactions between molecules - there are no three-molecule terms. Chemically, this assumes that the electronic structure of the molecule is only perturbatively affected by another nearby molecule.  It exploits the fact that quantum forces calculated via the Hellmann-Feynman theorem are simply electrostatic in nature. 
We further assume a rigid bond, i.e. that the \ce{N2}  vibration is in its ground state, such that it makes no spurious contribution to specific heat or entropy. 
These assumptions imply that our simple potential is designed to work at pressures where the triple-bond remains intact and at temperatures up to \qty{2000}{K}. 

Every intermolecular ($i-j$) configuration is uniquely described by the set of six interatomic distances defined between four atoms which are then used as an input to calculate atomic descriptors.
Here $i_1$ and $i_2$ are two bonded atoms of molecule $i$ and similarly $j_3$ and $j_4$ belong to molecule $j$.
The numerical subscripts label atoms in a given molecular interaction, such that the separation between atoms $i_2$ and $j_3$ is $r_{23}$.
In total, four atomic descriptors are computed, one per atom, for every molecule-molecule interaction.
Once all four descriptors are calculated, the energy for this particular interaction is obtained and contributions to the forces are integrated accordingly. The descriptors are then discarded and the process is repeated for next $i-j$ pair.

We use \textit{blip} basis functions, $B$, for the expansion of both two- and many-body descriptors \cite{Hernandez1997BasisCalculations}.
The \textit{blip} is composed piecewise out of B-spline polynomials in the four intervals [-2,-1], [-1,0], [0,1] and [1,2].
B-splines are localised basis functions used to represent functions in terms of cubic splines \cite{Schumaker2007SplineTheory}.
The \textit{blip} function is defined for our purpose as
\begin{equation} \label{eq::blips}
 B(r_b) =    
 \begin{cases}    
 1-\frac{3}{2}r_b^2+\frac{3}{4}|r_b|^3 & \text{if} \qquad 0<|r_b|<1\\    
 \frac{1}{4}(2-|r_b|)^3 & \text{if} \qquad 1<|r_b|<2\\   
 0 & \text{if} \qquad |r_b|>2    
\end{cases}                                               
\end{equation}
where $r_b=\eta(r-r_s)$ and $r_s$ is a parameter which centres the function on a grid position and $\eta$ controls its width such that $\eta/4$ is the span of a \textit{blip}.
The shape of the \textit{blip} functions is similar to Gaussians but because of their full localisation the number of computations can be significantly lower as the latter has infinite span. 
With the automated hyperparameter tuning, as implemented in \tadah, we were able to reduce both two- and many-body \textit{blip} grids to just four sets of parameters (SM).

The component of the pairwise descriptor of the $p$-th atom is accumulated by summing over three relevant distances using blip basis functions $B^n$ (eq. \ref{eq::blips}), where $n$ labels one of $\{r_s,\eta\}$ sets of hyperparameters.
\begin{equation}
v_p^{n} = \sum_{\substack{q=1 \\ q\neq p}}^4  B^n(r_{pq})f_c(r_{pq})
\label{eq::N2::blip_2b}
\end{equation}
where the sum runs over the neighbours of atom $p$ within its own molecule and in one adjacent molecule
and $f_c$ is the cosine function (eq. \ref{eq::cutoff::cos}) which ensures smooth energy cutoff
\begin{equation} \label{eq::cutoff::cos}
  f_c(r) =
    \begin{cases}
      0.5\Big[ \cos{\Big( \frac{\pi r}{r_c} + 1 \Big)} \Big] & \text{if $r \leq r_c$}\\
      0 & \text{otherwise}
    \end{cases}       
\end{equation}

The four-body interactions are captured by first computing local atomic densities using Gaussian Type Orbitals \cite{Zhang2018DeepMechanics,Takahashi2017} (eq. \ref{eq::N2::GTO}):
\begin{equation}
\psi_{l_x, l_y, l_z}^{\eta, r_s}(\mathbf{r}) = x^{l_x}y^{l_y}z^{l_z}\exp{\big(-\eta|r-r_s|^2\big)}
\label{eq::N2::GTO}
\end{equation}
where $x$, $y$ and $z$ are components of the displacement vector $\mathbf{r}_{pq}$ between two interacting atoms.

%Exponents $l_x$, $l_y$ and $l_z$ are the quantised directional-dependent angular momenta. The sum of those define the angular momentum $L=l_x+l_y+l_z$ in eq. \ref{eq::N2::MB} which effectively determine the order of the expansion. 

The summation is constrained (eq. \ref{eq::N2::MB}) to  ensure rotational invariance of the descriptor $\phi$, despite $\psi_{l_x, l_y, l_z}^{\eta, r_s}(\mathbf{r})$ not having this property.
\begin{equation}
    \phi^{L, \eta, r_s}_p = \sum_{l_x, l_y, l_z}^L \frac{L!}{l_x! l_y! l_z!}\Big(\sum_{\substack{q=1 \\ q\neq p}}^4 \psi_{l_x, l_y, l_z}^{\eta, r_s}(\mathbf{r}_{pq}) \Big)^2
\label{eq::N2::MB}
\end{equation}

In the current work, this many-body expansion is truncated at $L=1$, so that $l_x!, l_y!, l_z!$ and $L!$ are all equal to 1. Combined with the four choices of hyperparameters $\eta$ and $r_s$, this means we use eight components to the many body descriptor for each atom.
While in principle this results in a three-body descriptor (the expansion up to the p-orbital) it is found sufficient when combined with a linear regression and second order polynomial basis functions.
By taking the combinations of descriptor's components, an accurate representation of the four dimensional PES of two interacting \ce{N2} molecules is obtained.

The descriptor vectors are used to construct a design matrix $\boldsymbol{\Phi}$. The optimal set of weights, $\mathbf{w}$, is obtained by employing the Bayesian approach to linear regression 
\begin{equation}
    \mathbf{w} = \frac{1}{\sigma^2_{\epsilon}} \boldsymbol{\Sigma}_N \boldsymbol{\Phi}^T\mathbf{t}
\end{equation}
where, $\mathbf{t}$ is a vector of targets containing training energies and the covariance matrix $\boldsymbol{\Sigma}$ is given by
\begin{equation}
    \boldsymbol{\Sigma}_N^{-1} = \frac{1}{\sigma^2_p} \mathbf{I} + \frac{1}{\sigma^2_{\epsilon}} \boldsymbol{\Phi}^T \boldsymbol{\Phi}
\end{equation}

The complexity of the model is controlled by the ridge regression with the regularisation parameter $\lambda=\sigma^2_{\epsilon}/\sigma^2_p$ which is optimised by the evidence approximation algorithm \cite{Bishop2006}. This automated procedure avoids model overfitting given a sufficiently large training database.

\subsection{Training Database}

\label{N2::MLIP::Model1::TrainingDatabase}
The training set is built upon publicly available quantum chemistry data for two interacting \ce{N2} molecules \cite{Hellmann2013}.
Therein a coupled cluster method with single, double and noniterative triple excitations (CCSD(T)) were used to obtain 408 data points for 26 distinct angular configurations. The bond length was fixed at \qty{1.1014}{\angstrom}.
The data were further refined by including the effects of quadruple excitations, relativistic effects and core-core and core-valence correlations.
A five-site per molecule analytical model is given in the paper which allows one to generate CCSDT(Q)-based energies.
This model, called CCSDT(Q)-5, is the only external input to our training data - we do not use any DFT data.
In principle, CCSDT(Q)-5 could be applied directly in MD using 5 massless sites in the molecule; however, the presence of the Coulomb term results in the infinite span.
% , and a truncation function to reduce it to finite range. In practice, the additional cost of looping over 5 rather than 2 sites per molecule, and the slow iterative convergence of SHAKE for linear molecules proved more computationally expensive than using our model.

\section{Molecular dynamics}
Our \tadah package provides a plugin for the LAMMPS code to allow the custom made potential to be employed in large-scale MD.
The constraint on the bond lengths was enforced using LAMMPS fix SHAKE.  For a diatomic molecule this is trivial but for a five-site linear model such as CCSDT(Q)-5 the algorithm is unstable and can generate bogus dipole moments.

Calculations used the NPT ensemble with Nose-Hoover thermostat and barostat for single-phase materials, and the NPH ensemble for the two phase calculations. 
A timestep of \qty{1}{fs} is used throughout.

The melting curve is obtained using the phase-coexistence method with the following procedure. The initial box contains the relevant solid phase for a given pressure.
The initial configuration is equilibrated for \qty{20}{ps} in the NPT ensemble with temperature and pressure being close to the expected melt point.
After initial equilibration, approximately half of the box is kept frozen while the remaining molecules are first heated to $\qty{1.5}{T_m}$, where $T_m$ is experimental melting temperature, then cooled down to the initial temperature. The stages of melting and cooling takes \qty{10}{ps} each.
Finally the NPH ensemble is used to simulate the entire system for at least \qty{350}{ps}. The long simulation time is required for the system to equilibrate.
There are three possible scenarios at this stage. The molecules in the box either completely solidify, melt or a mixture of solid and liquid is present at the end of the simulation. The first two cases indicate that the initial temperature was too low or to high respectively. The latter case means that the simulation has equilibrated at thermodynamic pressure and temperature conditions somewhere on the melt curve.
The time averaged kinetic energy from the last \qty{50}{ps} is assumed to be corresponding to the melting temperature.

\section{NPT MD - Crystal phases} \label{SS:Crystal_Phases}
\begin{figure}[ht]
\includegraphics[width=0.45\textwidth]{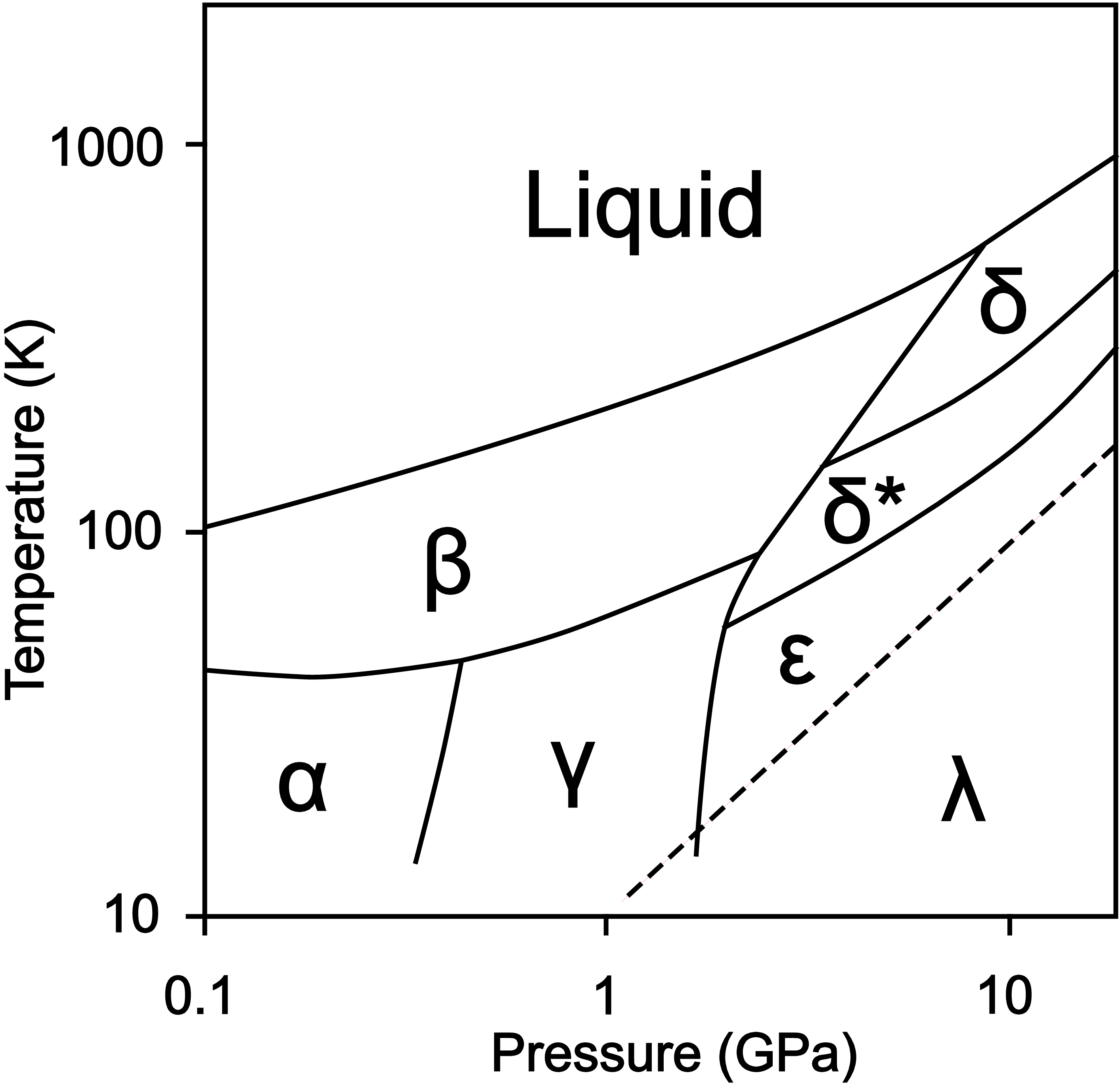}
\caption{\label{fig:Phase_Diagram_exp} The experimental phase diagram of \ce{N2} shows a number of competing phases. Figure adapted from \cite{gregoryanz2007high}, the \LAMBDA phase boundary (dashed line) from \cite{frost2016novel}, the melt curve is from \cite{Young1987}.}
\end{figure}

We begin exploration of the \ce{N2} system by running NPT MD at different pressure-temperature conditions to establish at least metastability of solid phases.
This also allows us to identify the approximate position of the melt curve and some of the solid phases boundaries. 
Those initial findings are then used to fully resolve the phase diagram of \ce{N2} (\ref{SS:Phase_Diagram}).
The obtained phase diagram is then compared with zero temperature DFT calculations (\ref{SS:DFT}).
The aim is to investigate all relevant phases under \qty{10}{GPa} which are shown on the experimental phase diagram in Fig. \ref{fig:Phase_Diagram_exp}.

\begin{figure*}
\includegraphics[width=0.95\textwidth]{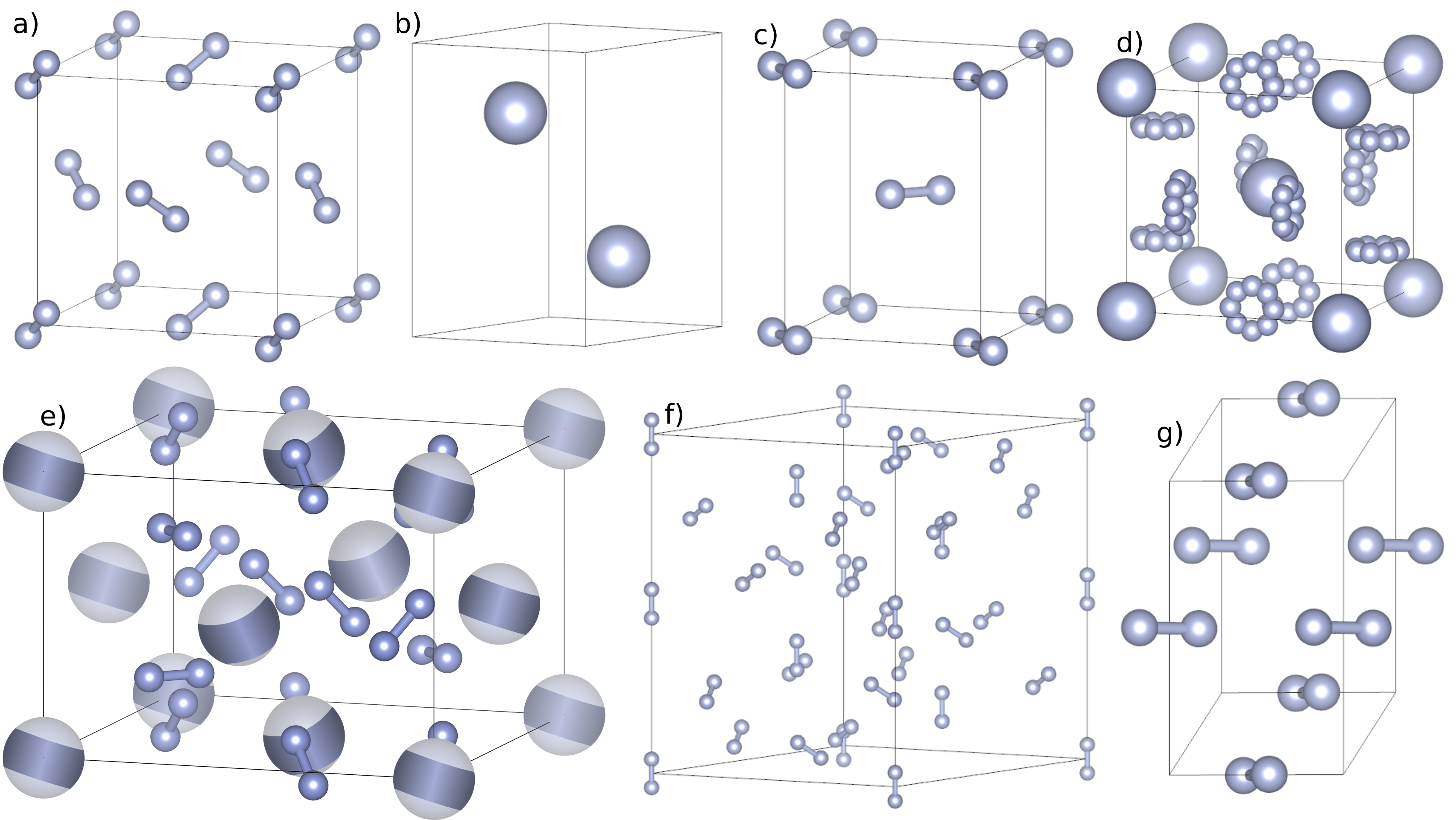}
    \caption{The solid phases of \ce{N2} under \qty{10}{GPa}. (a) \ALPHA, cubic with molecules along (111), (b) \BETA, hcp with rotating molecules, shown by spheres (c) \GAMMA, tetragonal, with molecules oriented along (110) (d) \DELTA, with spheres indicating 3D rotors and 8-atom rings indicating 2D rotors. (e) tetragonal \DELTALOC: the striped sphere show the preferred orientation of the molecules which primarily rotate about the tilted axis; the molecules are oriented in preferred direction, but reorient through 180 degrees on a picosecond timescale. (f) rhombohedral \EPSILON (g) monoclinic \LAMBDA.
    \label{fig:solid_phases}}
\end{figure*}

\subsection{Alpha phase (\ALPHA)}
Figure \ref{fig:solid_phases}a) shows alpha nitrogen \ALPHA which is a low temperature and low pressure phase.
The molecular centres of the \ALPHA are located on the face centred cubic (fcc) lattice. Each molecule is aligned along a different cube body diagonal which preserves cubic symmetry. 
In NPT molecular dynamics simulation the \ALPHA was found to be stable at temperatures above \qty{25}{K} which is in agreement with the experimental evidence \cite{Swenson1955,Schuch1970}. At temperatures below \qty{20}{K} it transitions to tetragonal \GAMMA phase. The computed lattice parameter at \qty{0.4}{GPa} and \qty{30}{K} is $a=\qty{5.42}{\angstrom}$ and is in excellent agreement with the experimental measurement \cite{Schuch1970}.
Upon heating, the \ALPHA libron oscillations increases and around the experimentally observed phase transition to the \BETA phase there is a sudden change to full 3D rotors. We note that the molecular centres remain on the fcc sites.

The compression of \ALPHA at \qty{15}{K} produces a transition to a twinned microstructure of \GAMMA at \qty{7}{GPa}, suggesting that the transformation path to this defective structure is martensitic. The transition is hindered by the high-energy barrier between the structures along a path that requires both unit cell strain and molecular rotation.  We use the NPT ensemble, but the twinning reduces the overall strain in the supercell.
The twin boundary has a higher energy than the perfect crystalline \GAMMA, indicating significant hysteresis in the transition.

\subsection{Beta phase (\BETA)}
The \BETA is a dominant high temperature phase up to around \qty{9}{GPa} with molecular centres located in a $P6_3/mmc$
structure close to hexagonal close packing (hcp).
The high symmetry $P6_3/mmc$ can be maintained if the molecules point along the $z$-axis, but this is implausible for a high-T phase and more likely indicates that the atomic positions are highly disordered or rotating \cite{Schuch1970} (see Fig. \ref{fig:solid_phases}a)).
The MD simulations show that the \BETA phase  remains stable in the PT conditions where it is experimentally observed. As expected, the molecules are close to freely rotating.
The hcp lattice parameters and their respective $c/a$ ratios, as obtained from MD simulations at experimentally relevant pressures and temperatures, are close to the \textit{ideal} $\sqrt{8/3}$ ratio for hexagonal close-packed hard spheres.
For example, the calculated lattice parameters at \qty{0.5}{GPa} and \qty{50}{K} are $a=\qty{3.835}{\angstrom}$ and $c=\qty{6.269}{\angstrom}$.
The obtained values agrees well with the experimental findings of $a=\qty{3.861}{\angstrom}$ and $c=\qty{6.265}{\angstrom}$ \cite{Schuch1970}.
Upon heating in the NPT ensemble the \BETA melts, while the quenching results in a twinned but ordered structure with molecular centres remaining on the original hcp sites.

\subsection{Gamma phase (\GAMMA)}
The \GAMMA phase is a low temperature and moderate pressure ordered phase of nitrogen and is shown in Fig. \ref{fig:solid_phases}c). 
Its crystal structure has been determined by X-ray diffraction as tetragonal with two molecules per unit cell at Wyckoff position $4f$ of space group $P4_2/mnm$ \cite{Schuch1970}.
Equivalently, \GAMMA can be described as a body centred tetragonal (bct) lattice with a central molecule pointing along $(110)$ direction and the corner molecule pointing along $(\overline{1}10)$, orthogonal to the central one. 
In NPT simulations at temperatures below \qty{50}{K} the \GAMMA phase remains at least metastable across the wide pressure range from \qty{0.1}{GPa} up to approximately \qty{5}{GPa}. The calculated unit cell parameters are within \qty{1}{\%} of the experimental ones \cite{Schuch1970}.
The structure can be related to \ALPHA via the Bain path as follows. One places the molecular centers on an fcc lattice, and reorients the molecules from pointing along $\{111\}$ (\ALPHA) to along $(110)$,$(\overline{1}10)$, (\BETA) this breaks the cubic symmetry. The c/a ratio drops from $\sqrt{2}$ to 1.29.

\subsection{Delta phase (\DELTA)}
   \begin{figure}[h]
        \centering
        \includegraphics[width=0.47\linewidth]{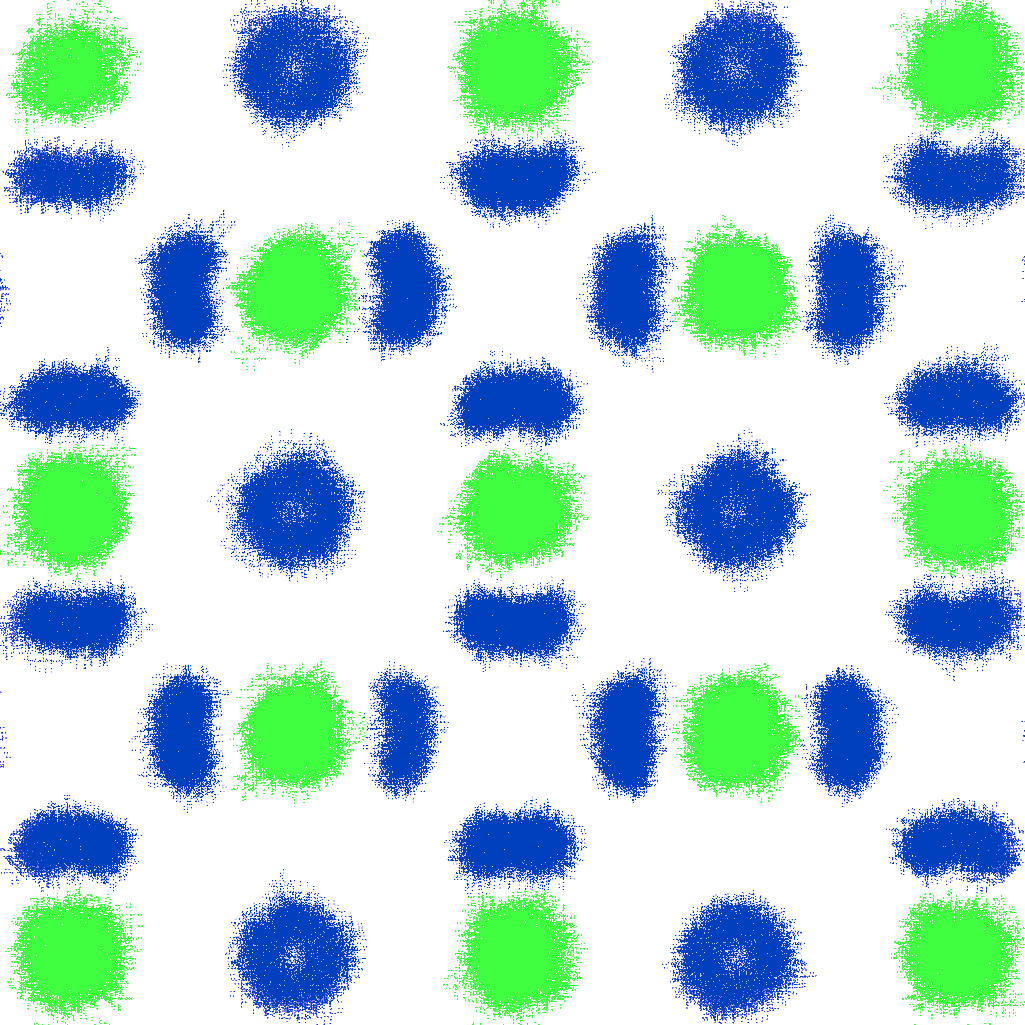}        
        \includegraphics[width=0.47\linewidth]{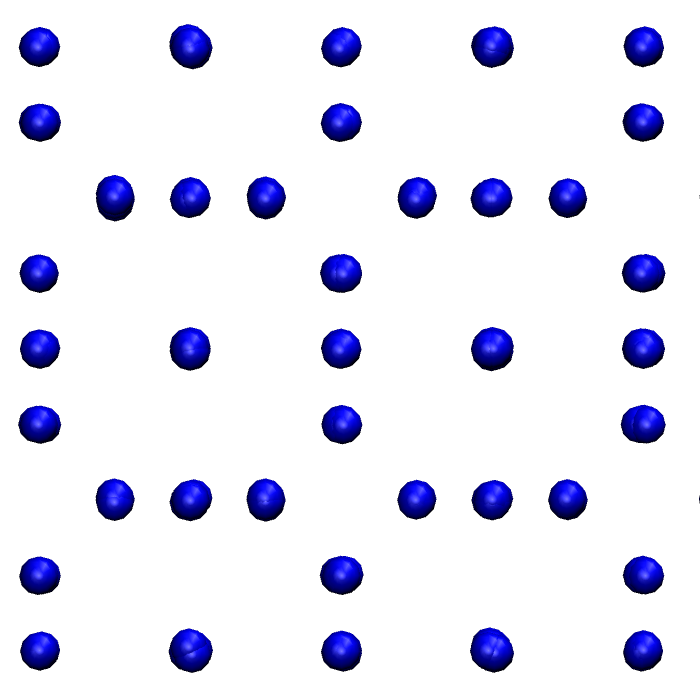}
    \caption{\label{fig:Delta_PDF}(Left) Probability density for atoms in the \DELTA phase from NPT MD at \qty{4}{GPa} and \qty{200}{K}. Two distinct molecular motions can be identified: sphere-like (green) and disc-like (blue). The kidney bean shape of discs indicates that the motion is not fully planar which is in agreement with the experimental findings \cite{Cromer1981}.
    (Right) Mean atomic positions averaged over \qty{1}{ps}. Molecules show nearly perfect spherical or disc-like motion.}
    \end{figure}
The cubic \DELTA phase has the Weaire-Phelan A15 structure ($\beta$-tungsten) space group $Pm\overline{3}n$ with eight molecules per unit cell and is similar to $\gamma-\ce{O2}$ and $\beta-\ce{F2}$ at \qty{50}{K} and atmospheric pressure  \cite{Cromer1981,mills1986structures}.
The unit cell of the \DELTA phase (Fig. \ref{fig:solid_phases}d)) consists of the two molecules located at $2(a)$ Wyckoff sites at $(0,0,0)$ and $(\frac{1}{2},\frac{1}{2},\frac{1}{2})$ which are approximately spherically disordered (represented as large spheres on Fig. \ref{fig:solid_phases}d)), however they preferentially avoid pointing along the cubic $\langle 100 \rangle$ directions \cite{stinton2009crystal}.
The remaining six molecules are located at $6(d)$ Wyckoff sites at $(0,\frac{1}{4},\frac{1}{4})$ and the respective cubic symmetry equivalents and their motion is disc-like.

The MD simulations confirms the existence of 3D rotors at $2(a)$ Wyckoff sites and discs on $6(d)$ sites (Fig. \ref{fig:Delta_PDF}).
It is observed that for both spheres and discs the center of the molecule moves significantly away from its respective symmetry site under thermal motion.
This results in a saddle shaped atomic distribution around the discs in agreement with \cite{Cromer1981}.
These disclike molecules form chains rotating around the x, y or z-direction. 
Heating of \DELTA does not result in solid-solid phase transitions: the phase ultimately melts once over the experimentally observed melting curve.
Quenching at \qty{5.7}{GPa} proceeds through initial ordering of the disc-like molecules and associated distortion of the unit cell to a tetragonal lattice.
The lattice parameter, as obtained from NPT MD, at \qty{5.7}{GPa} and \qty{293}{K} is $a=\qty{6.167}{\angstrom}$ agrees very well with the experimental value \cite{stinton2009crystal}.

\subsection{Delta* phase (\DELTALOC)}
\begin{figure*}
\includegraphics[width=0.95\textwidth]{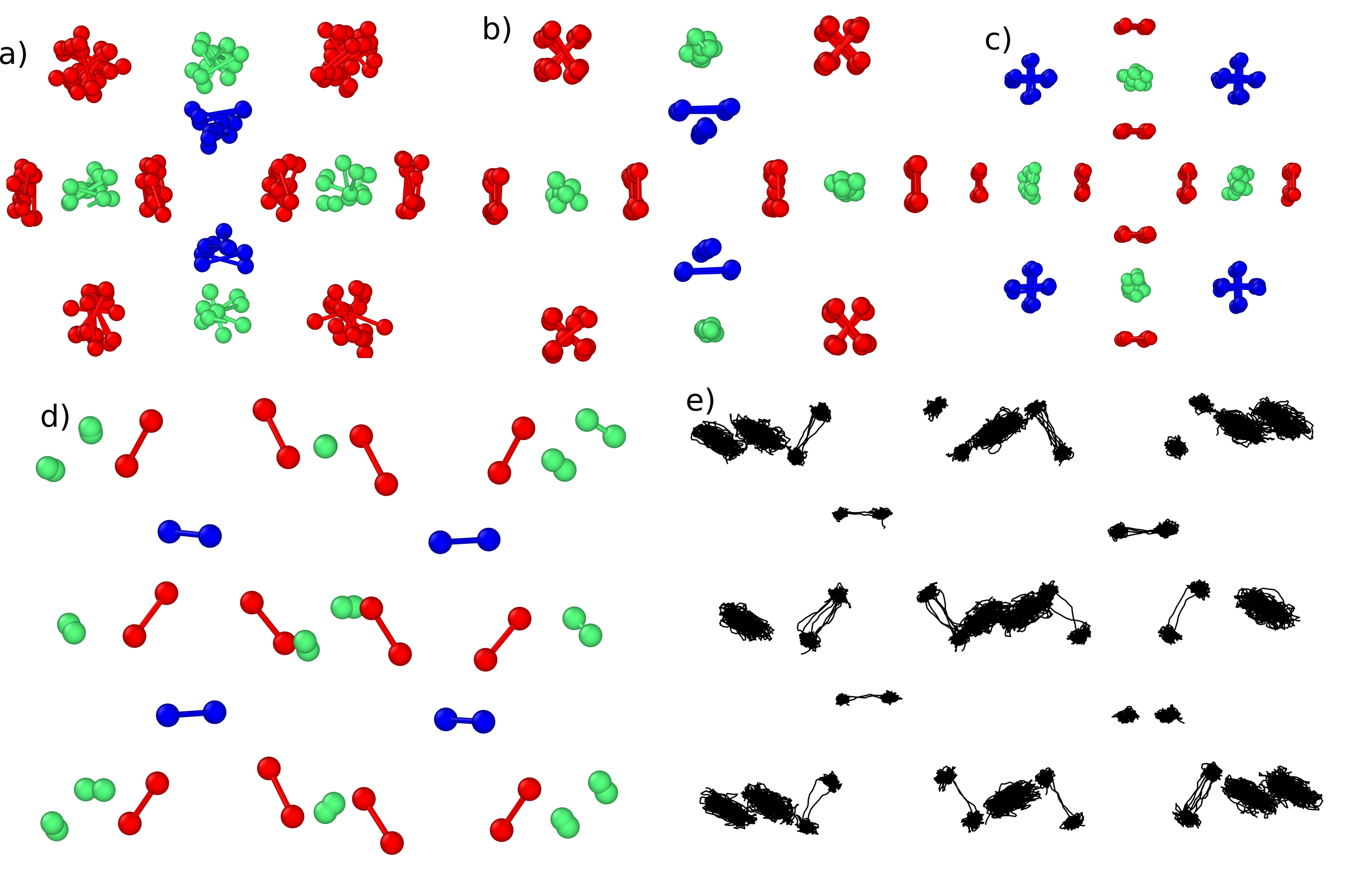}
    \caption{
    (a) MD snapshot taken from the NPT MD simulation of the \DELTALOC phase at \qty{120}{K} and \qty{5.7}{GPa}. The view is along [110]-direction.
    (b) and (c) The time averaged atomic positions over \qty{2.5}{ps} for [110] and [001] direction respectively.
    (d) The unit cell of the \DELTALOC phase obtained by time averaging atomic positions. The unit cell has the same orientation along the [010] direction as Fig. 7 in \cite{stinton2009crystal}. Green coloured molecules correspond to ex-spheres in \cite{stinton2009crystal} while red and blue to ex-discs of type 1 and 2 respectively.
    (e) The crystal orientation is as in (d) but with added trajectory lines and molecules removed for clarity. 
    The trajectory lines are computed over 4000 MD steps (\qty{200}{ps}) where atomic positions in every step are averaged using \qty{2.5}{ps} smoothing window.
    \label{fig:delta_loc}}
\end{figure*}

The tetragonal \DELTALOC (Fig. \ref{fig:solid_phases}e)) is a unit-cell doubling from \DELTA. Its space group has been proposed as $P4_2/ncm$ in 1998 by \cite{Hanfland1998} and finally resolved in 2009 by \cite{stinton2009crystal}.
The measured unit cell parameters are $a=\qty{8.063(5)}{\angstrom}$ and $c=\qty{5.685}{\angstrom}$ at \qty{14.5}{GPa} and \qty{293}{K} \cite{stinton2009crystal}, giving a c/a ratio just \qty{0.3}{\%} different from \DELTA.
The \DELTALOC phase is considered an intermediate phase between fully ordered \EPSILON phase and almost perfectly disordered \DELTA phase. The \DELTALOC shares the same positions for the molecular centres as \DELTA and \EPSILON \cite{stinton2009crystal,mills1986structures}.
However, in the \DELTALOC phase  all molecules appear to show preferred directions.
The refinement of the experimental structure has been performed in \cite{stinton2009crystal} who reported  disc-like coordinated motion where molecular orientations are either paired or perpendicular to each other.
However, they note that their proposed structure does not produce the observed number of Raman and infrared modes.

In MD simulations at \qty{10}{GPa} and \qty{200}{K} \DELTALOC simulation box is tetragonal with molecular centres remaining on $P4_2/ncm$ sites, however the rotations are significantly reduced relative to \DELTA.
The explanation for why this is beneficial, is as follows.
Along a line of ex-disc molecules which previously rotated about a z-axis (blue in Fig. \ref{fig:delta_loc}) the molecules now point alternately along (110) and $(\overline{1}10)$, with adjacent "chains" alternating (i.e. $(\overline{1}10)$ and (110)).  This freezing-in and period doubling alone would cause a cubic-tetragonal transition, to space group $P4_2/ncm$

The former-spherical rotors now rotate in 2D, about an axis with no obvious crystallographic direction, and its mirror image.  The rotating molecule is tilted away from the fixed direction of the ex-$z$-disc molecules
The tetragonal symmetry breaking means that the ex-discs rotation around $x$ and $y$ (red in Fig. \ref{fig:delta_loc}) remain symmetry equivalent.  Curiously, these lock into orientations roughly $(0, \frac{1}{2}, \pm \frac{\sqrt{3}}{2})$ with a fourfold AABB repeat. 
These molecules reorient through 180$^o$ on a picosecond timescale, significantly more often than the ex-$z$-discs.   But, they are essentially librating and there is no sign of static disorder.  All these observations are consistent with the X-ray data \cite{stinton2009crystal}. 

Thus, we can envisage the \DELTALOC phase as due to the 
 ex-$z$-disc molecules ceasing to rotate and forming a favourable ABAB chain with molecules at 90$^o$ to their neighbours.  This lock-in causes the ex-spheres to rotate preferentially about an axis which avoids the locked-in ex-$z$-disc.  Finally, the ex-$xy$-disc molecules also stop rotating, in directions so as to avoid the ex-spheres, which requires an AABB repeat.
 
The heating of \DELTALOC increases symmetry to cubic $Pm\overline{3}n$ structure (\DELTA) while cooling results in a distorted tetragonal lattice.
The high rate of quenching in MD simulations (approx. \qty{100}{K/ns}) results in a synthesis of crystals with a number of different molecular orientations, similar to \EPSILON but not always identical. We propose that there is a strong reduction in energy from orienting the molecules along one of several preferred directions, but only a weak additional energy gain from choosing the particular set of orientations associated with \EPSILON.  By comparison to the nanosecond timescale of the MD, in experimental settings involving disordered phases the sample is first annealed at high temperature before cooling it down at the slow rate of \qty{10}{K/hr} \cite{stinton2009crystal}.

At \qty{200}{K} and \qty{10}{GPa} the lattice parameters obtained from MD simulations are $a=\qty{8.36}{\angstrom}$ and $c=\qty{5.91}{\angstrom}$ giving c/a ratio of $0.706$ which is slightly below the $0.707$ expected from the cubic crystal. Such a small distortion from $Pm\overline{3}n$ to $P4_2/ncm$ is expected for this phase at relatively low pressure \cite{Hanfland1998} given the phase transition between \DELTA and \DELTALOC appears to be second order.

\subsection{Epsilon phase (\EPSILON)}
The  $R\overline{3}c$ rhombohedral \EPSILON phase is the  orientationally ordered version of the cubic \DELTA phase distorted along $\langle111\rangle$ direction - the resulting angle between axes is around $\qty{5}{\degree} $\cite{mills1986structures}.
The molecular center positions are slightly displaced as compared with the \DELTA.  The rhombohedral unit cell contains eight ordered molecules. \EPSILON remains stable at low T in approximately 2 to \qty{25}{GPa} range \cite{mills1986structures,Schiferl1985}.
The similarity between \DELTA, \DELTALOC and \EPSILON is apparent from their respective Raman stretching-mode spectra \cite{Schiferl1985} each containing two distinct  branches - intense lower frequency peak and less pronounced higher frequency peak, which can be associated with the 2(a) and 6(d) sites in \DELTA and their subsequent distortions.
The \EPSILON phase can either be obtained by compressing \GAMMA phase at low temperature or \DELTA phase at room temperature \cite{Olijnyk1990}. It is also possible to obtain it by slowly cooling \DELTA or \DELTALOC phases which results in ordering of molecules. However such an experiment is difficult to reproduce using limited timescale in MD simulations.
The MD hexagonal unit cell at \qty{10}{GPa} and \qty{80}{K} has dimensions $a=\qty{7.98}{\angstrom}$ and $c=\qty{11.07}{\angstrom}$. The unit cell and the increase in c/a ratio with pressure and agree well with the experimental values \cite{mills1986structures, Olijnyk1990}.

\subsection{Lambda phase (\LAMBDA)}
The \LAMBDA phase \cite{frost2016novel} has been suggested as having sheets of nitrogen molecules and $P2_1/c$ symmetry.  
A low energy DFT structure with $P2_1/c$ symmetry has been found with 4 atoms per unit cell, located on the 4e Wyckoff positions \cite{pickard2009high}. This phase is a good fit to the experimental X-ray pattern, but its two-molecule unit cell appears incompatible with the three vibrons observed in Raman spectroscopy.
The \LAMBDA phase can be considered as a distortion from the tetragonal \GAMMA phase.
The phase transition can be realised by gradually tilting \GAMMA molecules along the tetragonal c-direction. This implies a low (or zero) energy barrier on compression from the \GAMMA phase as there is no reshuffling of molecular centres.
The description of the structure as ``Layered'' is highly misleading: molecular centers lie close to an fcc lattice, with each molecule having 12 nearest neighbours.
 With our potential, the zero-temperature structure relaxation favours the \LAMBDA phase over \GAMMA, however the molecular rotation away from the (110)-direction is much smaller as compared with the DFT structure \cite{pickard2009high}. Our constant-stress NPT MD simulations in the experimental region where the \LAMBDA phase has been observed started with this initial structure, but  spontaneously transform the the \GAMMA phase.   Still, the \LAMBDA phase can be simulated with MLIP using an isobaric ensemble with fixed c/a ratio. 
 
 Experimentally, the transformations between ordered phases \GAMMA-\LAMBDA-\ALPHA have considerable hysteresis and are sensitive to details of the sample history.  Our calculations reflect this, showing that even at fixed temperature and pressure, one can drive the transitions by applied strain.

\subsection{Iota phase (\IOTA)}
The \IOTA phase\cite{turnbull2018unusually} was reported to be a 96-atom unit cell molecular nitrogen structure, with bondlengths ranging from 0.88 to 1.13\AA. The MLIP cannot describe this variation in bondlength, and DFT simulation gives an \IOTA structure with nearly equal bondlengths.   A simulation with the experimental cell and molecular orientations, but using the MLIP fixed bondlength, suggests that the \IOTA structure is metastable and has low energy, but is not the most stable phase.

\section{Validation via Ground state energies from DFT} \label{SS:DFT}
Although we do not use it for training here, Density Functional Theory (DFT) has become the standard method for producing data for machine-learned potential. 
Since no solid-phase data was used in the fitting,  the DFT ground state energy of the crystal structures is a good test of transferability.  
However, DFT is not a unique theory: results depend on the choice of exchange-correlation functional, and it does not give especially accurate results for systems with weak dispersion interactions.

\begin{figure*}
    \centering
   \includegraphics[width=1.0\textwidth]{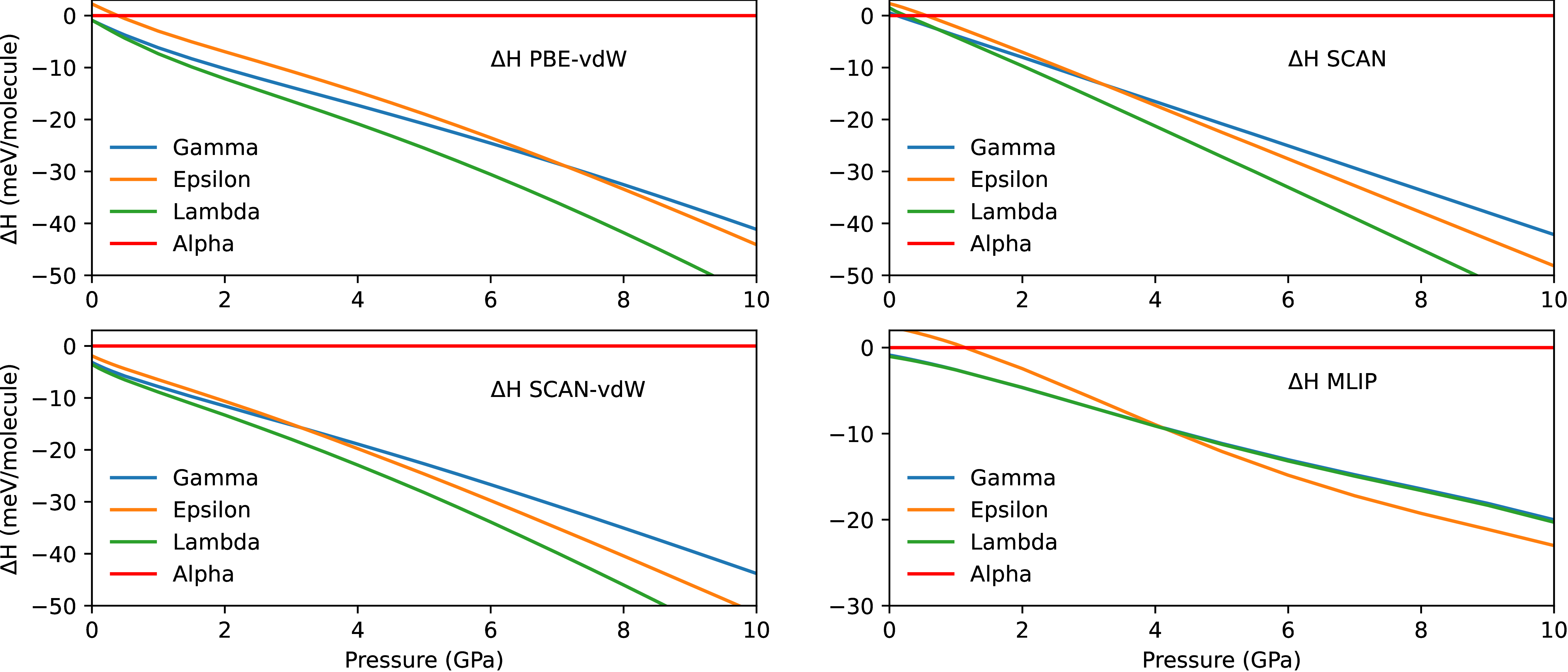}
   \caption{Enthalpy difference (relative to \ALPHA) between ordered \ALPHA, \GAMMA, \EPSILON and \LAMBDA phases as calculated by PBE-vdW, SCAN + SCAN-vdW, and the MLIP.  These calculations exclude phonon free energy. For exact transition pressures, see table \ref{tab:PT})}
   \label{fig:relH}
\end{figure*}

Unfortunately the situation with DFT is not clear: 
according to Materials Project \cite{MatProj}, the \ALPHA $Pa\overline{3}$ is unstable against a $P2_1\overline{3}$ distortion which was first proposed more than 60 years ago \cite{jordan1964single} but not proven in subsequent work \cite{schuch1970crystal}.  We calculated this structure, as well as the other ground states (\GAMMA, \EPSILON and \LAMBDA) with three different exchange-correlation functionals and MLIP.
In Fig. \ref{fig:relH} we see that the overall theory picture for is similar, with pressure favouring \LAMBDA over the remaining crystal structures. At very low P \ALPHA is the most stable phase for SCAN with the $\alpha\rightarrow\gamma$ transition pressure of \qty{0.1}{GPa}. However, the transition pressure is highly sensitive to choice of functional, with this transition even shifting to negative pressures for PBE-vdW, SCAN-vdW and MLIP.

Our potential also reproduces the correct sequence $\alpha\rightarrow\gamma\rightarrow\epsilon$ with increasing pressure, but like PBE it has \LAMBDA as the stable phase at zero pressure.  The MLIP predicts 
a transition to the \EPSILON phase at \qty{4.1}{GPa} which agrees with early measurements \cite{Schiferl1985, mills1986structures} but contrasts with recent experiment \cite{frost2016novel,turnbull2018unusually} where \LAMBDA or \IOTA is expected to be the most stable phase.

\begin{table}
    \centering
    \begin{tabular}{c|ccc} \\
         Method & $\alpha\rightarrow\gamma $ & $\gamma\rightarrow\epsilon$ &
         $\gamma\rightarrow\lambda$\\
         \hline \\ [-2ex]
        PBE-vdW & -0.1 & 7.1 & 0.0 \\
        SCAN & 0.1 & 3.2 & 0.7 \\
        SCAN-vdW & -0.4 & 3.1 & -0.1 \\
        Expt. & 0.35 & 1.9 & 2.0 \\
        MLIP & -0.6 & 4.1 & -\\
    \end{tabular}
    \caption{Transition pressures in GPa found using different methods. Experimental values from \cite{Schiferl1985, frost2016novel}. Negative transition pressures are estimated from data.}
    \label{tab:PT}
    \vspace{-10pt}
\end{table}

\begin{figure*}
 \centering
 \includegraphics[width=1.0\textwidth]{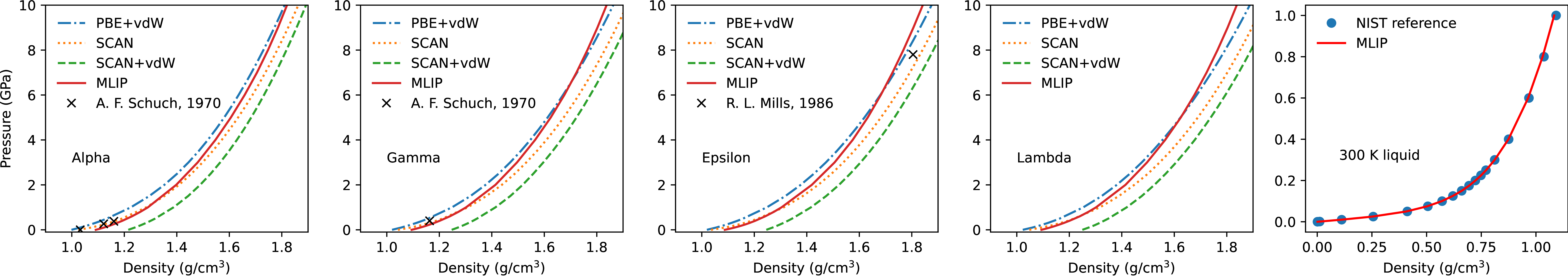}
  \caption{Equation of state for nitrogen calculated from DFT (PBE-vdW, SCAN, SCAN-vdW) and our potential at T=0. Rightmost  figure shows comparison of liquid T=\qty{300}{K} with the experimental reference \cite{Span2000}.}
   \label{fig:EoS}
\end{figure*}

This can be seen in Fig. \ref{fig:EoS} where different choices of functional give some 10\% variation in density at a fixed pressure.  Our potential falls within the uncertainty of these DFT calculations.  Similarly with the sequence of phase transformation under static relaxation (Fig. \ref{fig:relH}). 

All functionals show that the sequence with increasing pressure is $\alpha\rightarrow\gamma\rightarrow\lambda$, although the transition pressures are functional dependent.  In some cases the required pressure at T=0K is negative for the transition to happen: this is unphysical, but the implied densities can be reached through thermal expansion. However the enthalpy differences are very small, so thermal and zero-point effects may be significant.    

The \IOTA phase\cite{turnbull2018unusually} was reported to be a 96-atom unit cell molecular nitrogen structure which exhibits an exceptional range of intramolecular separations between 0.88 and \qty{1.13}{\angstrom}.  The MLIP cannot describe this variation in bondlength, and our DFT calculations, set up in the experimental structure, relax to equalise the bonds at a conventional length, around \qty{1.08}{\angstrom}.  These relaxed structures are metastable, but the wide range of bondlengths,  reported for this structure cannot be understood with either DFT or MLIP.
 
The MLIP gives similar qualitative behaviour, with EoS and transition pressures within the uncertainty of DFT functionals. For the present study, this is the best possible DFT-based validation of the MLIP.

\section{Face Centred Structures} \label{SS:FCC_Structures}
One could consider the three phases \ALPHA, \GAMMA and \LAMBDA as decorations of molecules on a face-centred cubic/tetragonal lattice.  In each case, the symmetry-breaking to $Pa\overline{3}$,  $P4_2/mnm$, and $P2_1/c$ is fully determined by the orientation of the diatomic molecule centred on an fcc lattice. The relaxation of the lattice to tetragonal or monoclinic does not introduce any further change in symmetry. 

The fact that these lattice relaxations are small is concealed by the choice of unit cell reported.  In fact, the tetragonal distortion of the \GAMMA phase is only 3\% in $a$ and $b$, and 6\% in $c$ away from cubic.  In \LAMBDA at \qty{5}{GPa} our SCAN-DFT calculations give $a=\qty{3.473}{\angstrom}$, $b=\qty{3.481}{\angstrom}$, $c=\qty{6.354}{\angstrom}$, $\beta = \qty{132}{\degree}$. Compared to conventional fcc, these axes map to $(\frac{1}{2},\frac{1}{2},0)$, $(\frac{1}{2},-\frac{1}{2},0)$ and $(-\frac{1}{2},-\frac{1}{2},1)$, giving a distortion from fcc of 1.3\%, 1.5\% and \qty{7}{\degree}.

These phases are thereby linked through martensitic transformations.  If post-transformation structures are examined via single-crystal diffraction, strain relaxation will result in twinned microstructures\cite{bhattacharya}.  We have seen this in simulations, where the transformations in NPT molecular dynamics give twinning, but the transition can be realised by applying strain (varying the c/a ratio). 

These structures have molecules oriented in different directions,
and can be mapped to an antiferromagnetic fcc lattice, the decoration of which remains a contentious issue \cite{ehteshami2020phase,ackland2023existence}.  Quadrupole interactions do not favour alignment, and are reasonably long ranged.  The Pa$\overline{3}$ arrangement is the most favourable decoration of an fcc lattice with quadrupoles, but it is unstable against lattice distortions\cite{van2020quadrupole}.   In early work\cite{NoseKlein1986}, it was thought that \BETA might also be fcc with rotating molecules, and indeed such a structure can be generated in MD by heating \ALPHA.     In fact \BETA is based on hcp, but the simulations show that the molecular orientations become (dis)ordered far faster than the molecular centers can rearrange.  So, heating transformations pass through a metastable intermediate fcc rotor phase, while cooling transformations pass through a metastable intermediate state of quadrupole ordering in hcp, which has multiple competing states and is very prone to domain formation\cite{van2020quadrupole}.   Such domains are likely to result in heavily twinned crystals, and complex diffraction patterns which can be challenging to solve.

\section{\ce{N2} Phase Diagram from MD} \label{SS:Phase_Diagram}

We build the phase diagram in stages.  Firstly we calculate the melt curve for both hcp \BETA and cubic \DELTA phases as well as a hypothetical fcc phase with 3D rotors.  The crystal phase with the higher melting temperature has lowest free energy, and  is stable, and the intercept is the  $\beta$ - $\delta$ - liquid triple point.

Since we know the latent heat and the density difference between the phases, we can also use the Clausius-Clapeyron equation to determine the initial direction of the \BETA and \DELTA phase line away from the triple point.

Along the zero-Kelvin line, we can use enthalpy calculations to locate the $\alpha$ - $\gamma$ -$\epsilon$ intercepts, and the Third Law to determine that the initial phase boundary is vertical.  We also run $N\dot{P}T$ simulations across the $\alpha-\gamma$ phase boundary which allows the Bain transition to be mapped, albeit with hysteresis.

Several of the phase transformations  involve a transition from fixed orientation to rotation of the molecules.  We can model this with $NP\dot{T}$ simulations, gradually heating the sample until the rotation starts.  The phase line can be crossed in both directions. 

The ordered crystal phases can be unambiguously identified by time averaging NPT trajectory and analysing obtained crystal with symmetry analysis package such as spglib \cite{spglib2018}. For the rotor phases we repeat the procedure using molecular centres of mass, and use probability density plots (e.g. Fig.\ref{fig:Delta_PDF}) to determine rotations.

\subsection{The Melting Curve} \label{SS:Melting_Curve}
\begin{figure}[ht]
\centering
\includegraphics[width=0.45\textwidth]{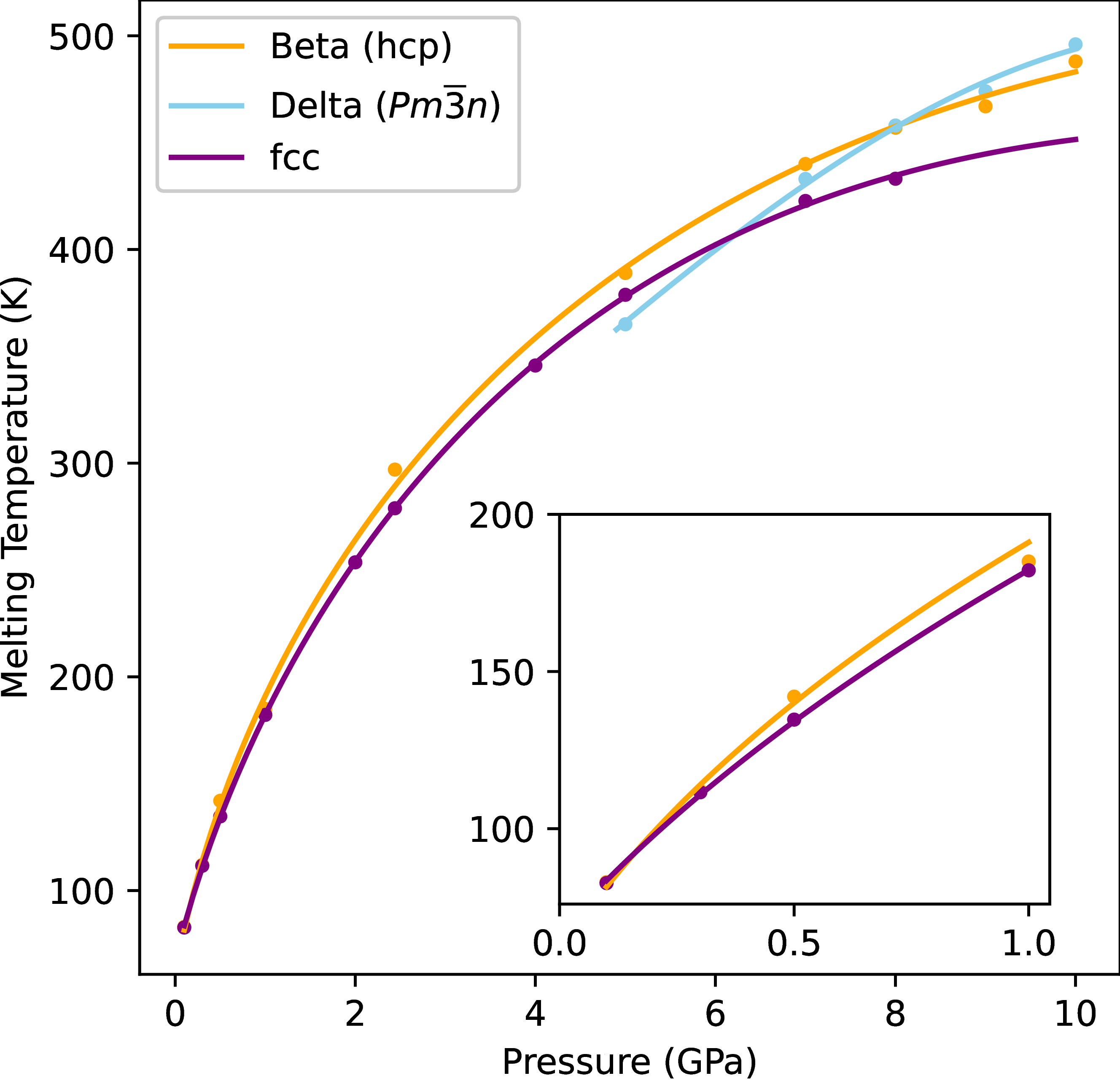}
\caption{\label{fig:Melt_Curve}The melting curves obtained using two phase coexistence method for experimental \BETA and \DELTA phases compared with hypothetical fcc structure with rotors. Higher melting point for a given pressure indicates energetically favourable structure. The curves are fitted using modified Simon–Glatzel equation \cite{Kechin_1995}.}
\end{figure}

We used phase coexistence calculations to track the melt curve of the relevant phases. The total of 25 coexistence calculations at 10 different pressure points were performed to obtain smooth melt lines as shown in Fig. \ref{fig:Melt_Curve}.

The simulation box for both the \BETA phase and the hypothetical fcc phase with rotors contains 28800 molecules on the hcp and fcc lattices respectively. The molecular orientations are initially assigned at random and show full spherical disorder/rotation throughout the simulation.

The initial structure for \DELTA has space group $Pm\overline{3}n$. The cell is constructed with 29952 molecules. The molecular orientations where randomly assigned: disclike for  $6d$ Wyckoff and spherical disordered for $2a$.

The calculated melting temperatures under various pressures are shown in table \ref{tab:phase_coexistence}.
The obtained melt curve is in a very good agreement with the experimental one from Fig. \ref{fig:Phase_Diagram_exp}.

In addition to determining the melt curve, these calculations pinpoint the triple point: it is at the pressure at which \BETA and \DELTA have the same melting point, \qty{8}{GPa} and \qty{457}{K}. The fcc structure with rotors has lower melting temperature and is therefore energetically unfavourable across the measured pressure range.

\begin{table}
\vspace{5pt}
\begin{tabular}{c|c|c} 
 P (GPa) & $\beta$ T (K) & $\delta$ T (K) \\
 \hline
 0.1 & 83 & - \\ 
 0.3 & 112 & - \\
 0.5 & 142 & - \\
 1.0 & 185 & - \\
 2.45 & 297 & - \\
 5.0 & 389 & 365 \\
 7.0 & 440 & 433 \\
 8.0 & 457 & 458 \\
 9.0 & 467 & 474 \\
 10.0 & 488 & 496 \\
\end{tabular}
\caption{The melting temperatures obtained from the phase coexistence simulations between \qty{0.1}{GPa} and \qty{10.0}{GPa} for both $\beta$ and $\delta$ phases of nitrogen. \label{tab:phase_coexistence}} 
\vspace{-5pt}
\end{table}

\subsection{Solid-solid Phase Boundaries}

\subsubsection*{alpha/gamma $\longleftrightarrow$ beta}
\begin{figure}
\centering
\includegraphics[width=0.9\columnwidth]{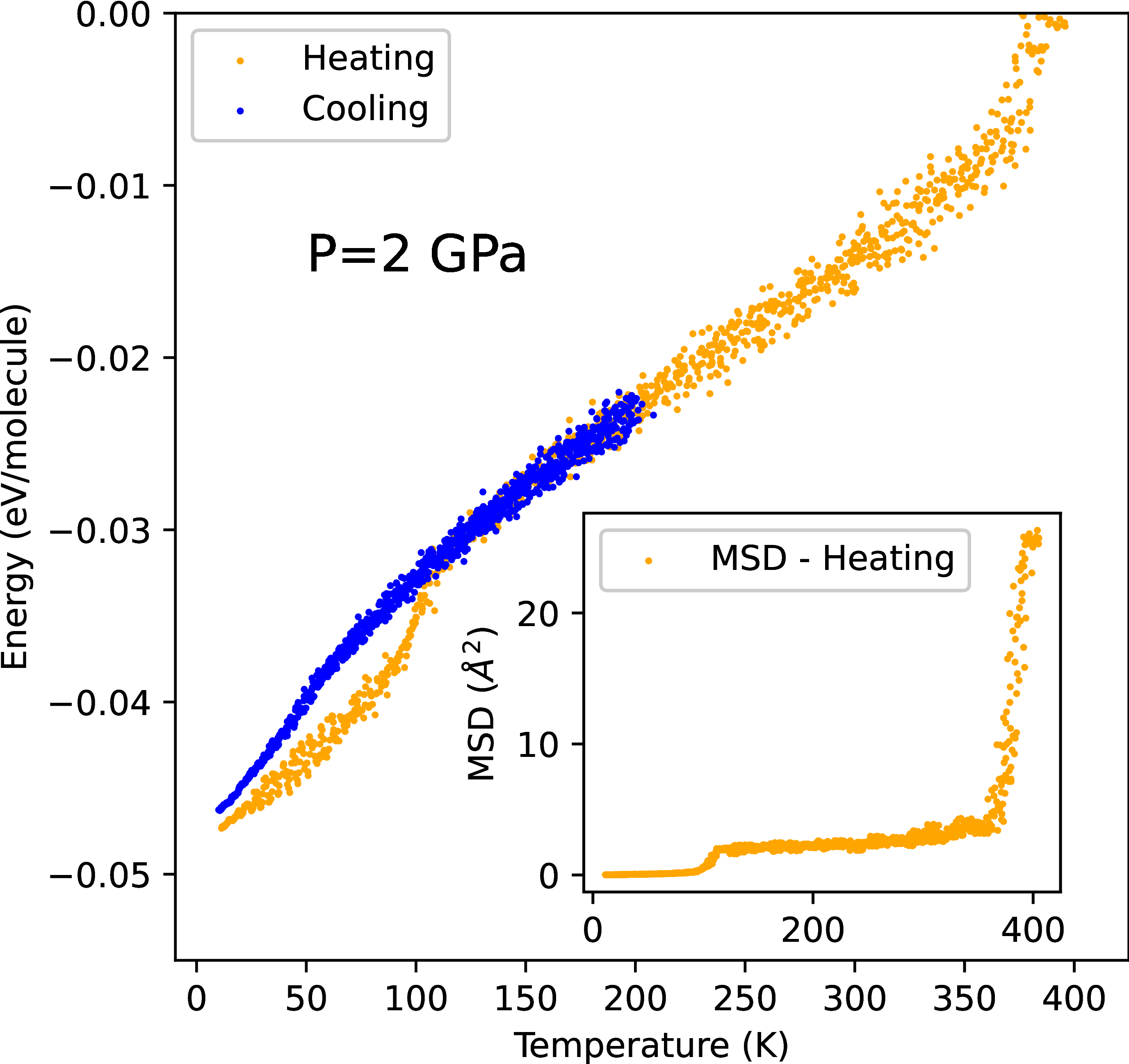}
\caption{Mean potential energy vs temperature from a single MLIP run starting in  $\gamma$ at P=\qty{1}{GPa} and heating rapidly from \qty{10}{K} to \qty{300}{K} through the sequence $\gamma\rightarrow\beta\rightarrow$melt. The inset shows corresponding mean square displacement obtained during heating. Three distinct stages can be identified: molecular solid with librons, solid with rotors and finally melt.
\label{fig:Melting_Gamma}}
\end{figure}
The phase boundary between $\alpha-\beta$ and $\gamma-\beta$ is represented on the phase diagram in Fig. \ref{fig:Phase_Diagram_MD} as a solid black line with an increasing positive slope with pressure. The phase boundary has been estimated based on the following observations.

For the $\alpha/\gamma\rightarrow\beta$ transformation, the molecular centers are completely different, and the transition is sluggish such that on heating we see a transformation from ordered fcc \ALPHA to a metastable free rotor fcc.  Alternatively, on cooling from rotating hcp \BETA , we obtain a metastable ordered hcp.  These two transitions bound the true transition line.
Near the experimental boundary it is observed that during the molecular dynamics simulations across the boundary, molecules in the \ALPHA phase begin to rotate or equivalently \BETA rotors cease their motion. 
This behaviour is only observed in the narrow temperature zone, and we take it as indicating the position of the phase boundary between librating and rotating molecules. If we consider the static hcp and rotating fcc phases to be metastable, then the \ALPHA heating simulations give an upper bound on the true phase line, while the \BETA cooling calculations give a lower bound: these are close enough to determine  the phase boundary with small errors. 

Although a rotation$\leftrightarrow$libration transition is observed in both fcc (\ALPHA)  and hcp (\BETA) on heating and on cooling, we note that no direct phase transition is obtained from fcc to hcp or vice versa in any simulation.  
The molecular centers in simulations started in the \BETA phase remain hexagonal even after rotation ceases;  similarly the molecular centres of \ALPHA remain close to fcc.  In both cases, the rotation ceases at similar T,P conditions.  

The fact that the transition does not occur on a nanosecond timescale is consistent with the experimental observation that the transformation is sluggish. This is understandable as even for an atomic system fcc/hcp phase transition is complex and difficult to realise in molecular dynamics simulation.
For example, the hcp to fcc transition in titanium is a process which involves slip of planes dislocations, adjustment of interplanar spacing followed by the volume expansion \cite{Yang2018}.
To best of our knowledge the mechanism of the fcc/hcp phase transition for \ce{N2} (or any other) dimers is unknown.

\subsubsection*{beta $\longleftrightarrow$ gamma}
\begin{figure}[ht]
\centering
\includegraphics[width=0.50\columnwidth]{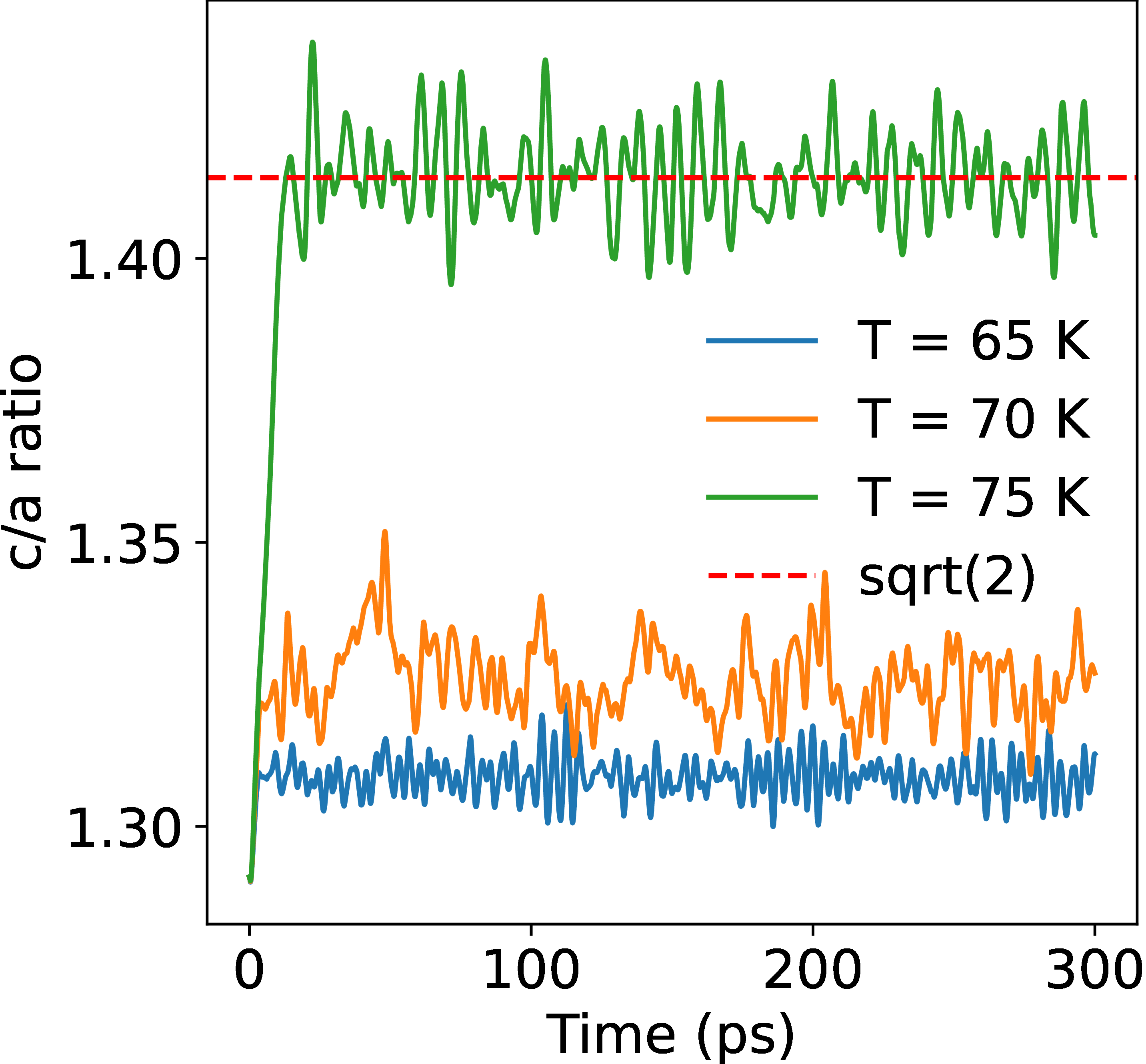}
\includegraphics[width=0.46\columnwidth]{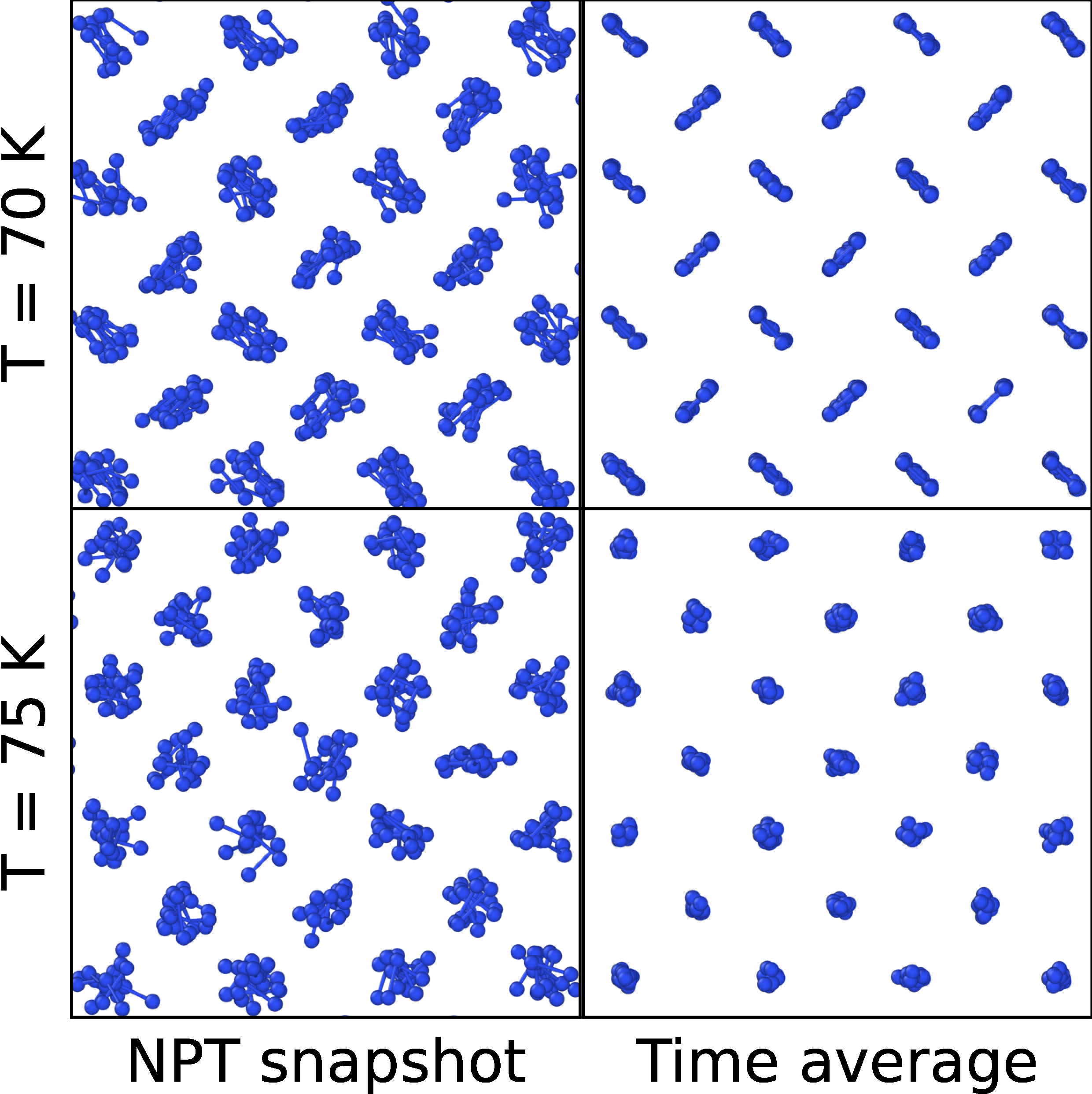}
\caption{\label{fig:Bain_path}(Left) Figure shows the c/a ratio of the $\gamma$-N2 phase at various temperatures simulated in the NPT ensemble at \qty{1}{GPa}. At \qty{65}{K} and \qty{70}{K} librons are observed with occasional disc-like motion fixed to either $110$ or $\overline{1}10$ rotation axis. At \qty{75}{K} there is a sudden increase in c/a ratio associated with molecules transitioning to free rotors. After the phase transition to molecular centres are located at fcc sites. (Right) MD snapshots showing the c-direction at \qty{70}{K} and \qty{75}{K} and respective \qty{2}{ps} time averages.}
\end{figure}

Fig. \ref{fig:Bain_path} shows that upon heating of the tetragonal \GAMMA phase, there is a small increase in the $c/a$ ratio. Close to the phase boundary the molecules begin to rotate and there is a sudden jump in the $c/a$ ratio to $\sqrt{2}$ indicating a transition to a perfect fcc lattice - the same as obtained from heating alpha.  
The initial transition from body centered tetragonal lattice to fcc follows classical Bain transformation \cite{Bain1924}. Therefore we speculate that phase transition from both \ALPHA and \GAMMA to the \BETA phase proceeds through metastable rotationally disordered fcc phase.
The complete phase transition to the hexagonal \BETA phase does not occur due to the same reason as for $\alpha\rightarrow\beta$ as explained above. 

\subsubsection*{epsilon $\longleftrightarrow$ delta$^*$}

\begin{figure}
\centering
\includegraphics[width=0.9\columnwidth]{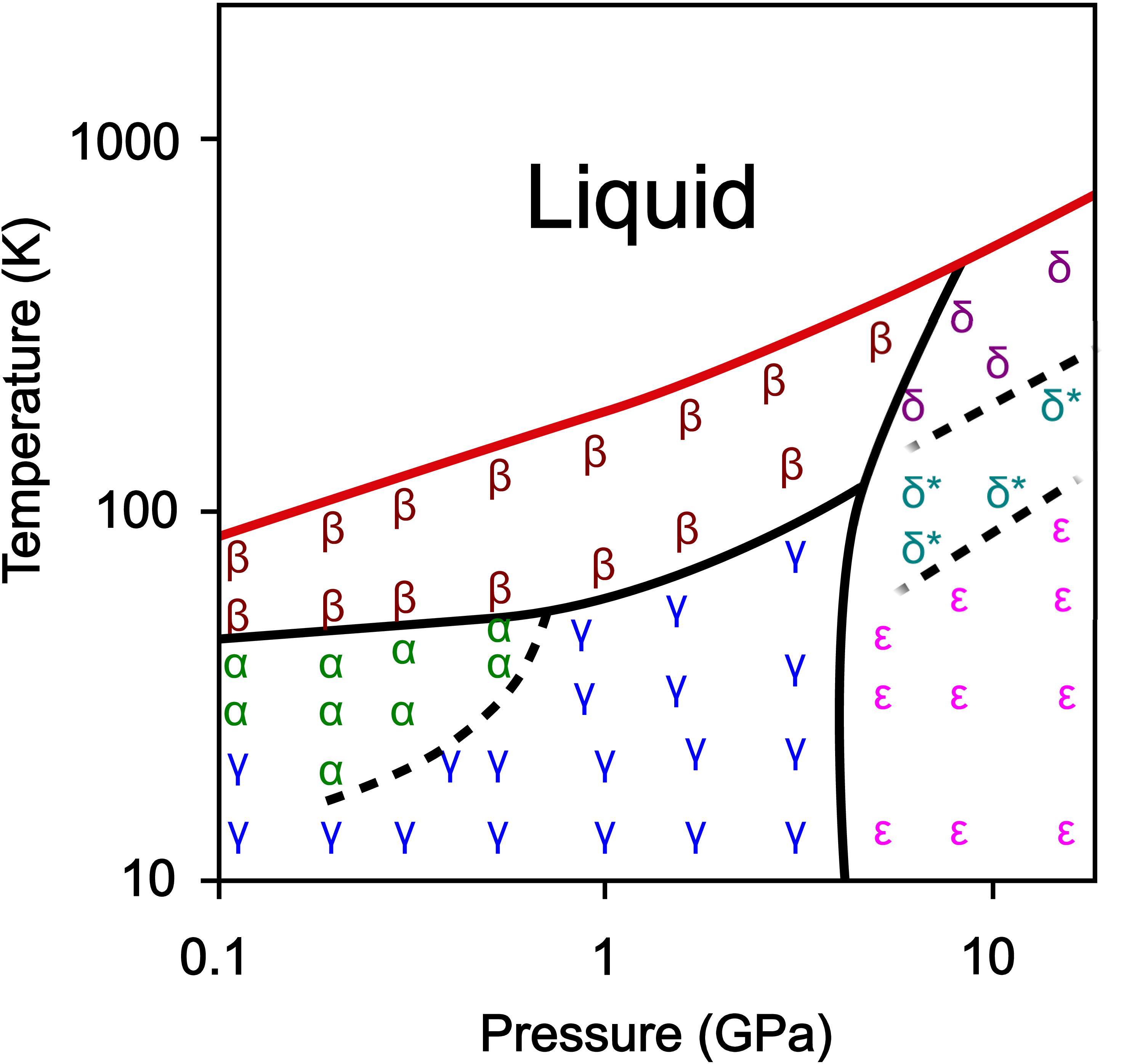}
\caption{\label{fig:Phase_Diagram_MD} The phase diagram for solid nitrogen obtained using MLIP potential.  Symbols show structures observed in MD at the given P,T conditions, lines are derived from coexistence and free energy calculations as described in the text.
}
\end{figure}

On heating from \EPSILON at \qty{8}{GPa} it is observed that the molecular motion changes and molecular orientations progressively evolve from libron-like motion to spherical rotor or disc-like motion. $NP\dot{T}$ ensemble heating runs along several isobars were performed to determine the approximate position of the line. 
Additional $NPT$ simulations were performed close to the phase boundary, and crystal symmetry analysis utilised to identify transition points from rhombohedral $R\overline{3}c$ to tetragonal $P4_2/ncm$ of the \DELTALOC phase.
The dashed line in Fig. \ref{fig:Phase_Diagram_MD} separating \DELTALOC and \EPSILON phases indicates the position of the established phase transition.

\subsubsection*{delta $\longleftrightarrow$ delta$^*$}
We were able to establish boundaries between \DELTALOC and  \DELTA phases by time averaging NPT simulations over \qty{1}{ns} after initial equilibration. 
We notice cubic $Pm\overline{3}n$ to tetragonal $P4_2/ncm$ phase transition when quenching the \DELTA phase. This observation is further confirmed by heating \DELTALOC. We do not observe any significant hysteresis which indicates very low energy barrier between two phases. This is perhaps unsurprising given that \DELTALOC phase is just a small distortion from the cubic \DELTA phase.

%The transformation from fixed orientation \EPSILON to rotations is more gradual in the A15-based structures epsilon-delta than in the close packed alpha-beta types, and we speculate that the reported $\delta^*$ phase comes from intermediate structures between epsilon-delta with significant short-ranged order and preferred orientation.

\subsubsection*{beta $\longleftrightarrow$ delta/delta$^*$}
The intercept of this line with the melt was determined from the phase coexistence calculations.
Similarities between crystal structures of \DELTA, \DELTALOC and \EPSILON phases as well as zero temperature phase transition between \GAMMA and \EPSILON allow us to trace the \BETA to \DELTA/\DELTALOC line.

\subsubsection*{alpha $\longleftrightarrow$ gamma}

Static relaxation (Fig. \ref{fig:relH}) shows that the \GAMMA phase has lower enthalpy than \ALPHA at \qty{0}{K} and all pressures.  This surprising result turns out to be consistent with DFT calculations.  Nevertheless, it is clear in both DFT and MLIP calculations that alpha becomes stable at densities only marginally larger that those calculated at T=0, P=0. These densities are obtained by thermal expansion at finite temperature (and through consideration of zero-point energy). Therefore this surprising finding regarding theoretical enthalpy does not contradict experiment.

It is observed that during MD simulations the \ALPHA phase spontaneously transforms to \GAMMA at temperatures below \qty{20}{K} across the computed pressure range. The phase transition occurs within the first few picoseconds of the simulation.
This observation is further confirmed by comparing MD enthalpy differences at pressures below \qty{0.5}{GPa} as shown in Fig. \ref{fig:relH}.
The \GAMMA phase is marginally more stable at low pressure but its relative stability increases suddenly near the expected phase boundary at \qty{0.3}{GPa}. 
The obtained free energy differences at zero temperature are around \qty{1}{meV/molecule} in the measured pressure range.
The small enthalpy differences and directly-observed phase transition indicates low phase transition energy barrier below \qty{20}{K}.
However, above those temperatures the phase transition is not observed indicating at least strong metastability of the \ALPHA phase.

%The experimental evidence points to \ALPHA being stable at low temperature and low pressure.The behavior of the model is therefore different with the \GAMMA phase being the most stable low temperature phase.

\subsubsection*{gamma $\longleftrightarrow$ epsilon}
The intercept of the \GAMMA-\EPSILON line with the P axis was precisely determine by static relaxation calculation. The Third Law of thermodynamics means that line must be vertical at T=0. The transition was not directly observed in MD, so we simply interpolate between the known points.

\section{Discussion and conclusions}
The success of our model provides strong evidence that our chemically-understandable descriptors are sufficient to capture the necessary physics, whereas conventional rigid molecules with pairwise atom-atom interactions, quadrupoles, or formal charges have failed.  The relative simplicity of our model means that it provides a good agreement with the experimental data for all molecular phases of nitrogen except the \IOTA phase. By enabling molecular dynamics, we are able to describe the nature of each phase in more detail than is possible experimentally.

We deduce that \ALPHA Pa$\overline{3}$ is favoured by quadrupole-quadrupole interactions, which are captured by our four-body interatomic descriptors.
The small distortion of Pa$\overline{3}$ to P2$_1\overline{3}$ is washed out by thermal effects.

The \GAMMA phase is a slightly more efficient packing than \ALPHA.  In the transformation, molecules rotate away from $<111>$ so as to point along $<110>$ directions.  This breaks cubic symmetry, leading to  martensitic-type transformation as the c-axis contracts. 

The \BETA phase is revealed as freely rotating \ce{N2} molecules in hexagonal close packing.  The transformation between \ALPHA and \BETA involves two aspects, the rotation of the molecules and the transformation from fcc-like to hcp-like crystal.  The onset of rotation occurs rapidly with heating, but the fcc-hcp transformation is sluggish and prone to generation of stacking faults\cite{loach2017stacking}.  In molecular dynamics, the $\alpha\rightarrow\beta$ transformation on heating goes first to a rotating fcc phase. 

Regarding the transitions, $\alpha\rightarrow\beta$, $\gamma\rightarrow\beta$ and $\epsilon\rightarrow\delta$ are revealed as due to the extra entropy available to molecular rotors.  Consequently, we expect these transformations to be temperature-driven  with flattish phase boundaries.  By contrast $\alpha\rightarrow\gamma$, $\gamma\rightarrow\epsilon$, is primarily a competition between energy and density, so it is pressure driven leading to nearly vertical phase lines.  

The $\beta\rightarrow\delta$ transformation combines both aspects of the transition.  Converting 3D rotors to 2D rotors comes at an entropy cost, but there is a density increase in packing the non-spherical objects.  Consequently, the $\beta\leftrightarrow\delta$ transition line has a positive slope.  Similarly, the liquid has higher entropy and lower density than coexisting solid phases, so the melt curve also has a positive slope. 

The $\alpha\rightarrow\gamma$ transition is driven by enthalpy, with the higher-density gamma phase observed at higher pressure.  However, the MD also suggests there is a significant entropy difference in favour of $\alpha$, so the transition line has a positive slope.  Extrapolating the transition line from classical MD to T=0 suggests that \GAMMA is the T=0, P=0  ground state, a surprising result which is consistent with DFT calculations. 
However the enthalpy difference is small.

The MD sheds considerable light on the nature of the $\delta$ and $\delta^*$ phases.  Both are based on the topologically close packed Weaire-Phelan A15 crystal structure ($Pm\bar{3}n$) and exhibit dynamic disorder with 3D and 2D rotors.  In \DELTA there is a body centred-cubic sublattice with 3D rotors, with the remaining 3/4 of the molecules in the cube faces being 2D rotors with the axis in a (001)-type direction.  \DELTALOC has molecules centred on the sites, with the bcc sublattice showing 2D rotation, and the cube-face molecules being non-rotating 
 (possibly disordered on rapid cooling).  When all rotation stops, the \EPSILON phase is formed.

The model supports the existence of the \LAMBDA phase at low temperatures.  Regarding \IOTA, although the reported spread of bondlengths is implausible, we find that a structure with equalised bondlengths and the same pattern of orientations is competitive.

On a broader view, we find a strong correlation between the unit cell shape and the molecular orientations: rather small strains in the cell can result in significant changes in orientation.  At a mesoscopic level this leads to microtwinning, but we cannot rule out \IOTA-like complex structures forming in response to non-hydrostatic strain, and in doing so relaxing the non-hydrostatic stress.

In many cases, single-molecule reorientations have low energy and there could be significant numbers of them thermally-activated.  We can speculate that these populations of defects may produce well-defined signals in spectroscopy, perhaps resolving the discrepancy between Raman and X-ray/DFT for  \LAMBDA. 

Finally, we emphasize that the insights gained into solid nitrogen should not overshadow the transferrability of the MLIP itself.   We believe that this is the first MLIP for a molecular material capable of describing such a wide range of crystal structures, none of which were include in the training data. Its simple form provides a deep physical understanding that, indeed, the properties of solid \ce{N2} are fully determined from the interactions between pairs of molecules.

%and quadrupole moment of the nitrogen molecules decreases with pressure as the crystal environment becomes denser \cite{LeSar1984ImprovedGasModelN2}. This might indicate that our model underestimates quadrupole moment causing gamma to be most stable at zero P. 

%\section{Conclusions}

\begin{acknowledgments}
The authors would like to acknowledge the support of the European Research Council (ERC) Grant ``Hecate'' reference No. 695527. MK and PIC acknowledge support from EPSRC for studentships, 
We are grateful for computational support from the UK national high performance computing service, ARCHER2, for which access was obtained via the UKCP consortium and funded by EPSRC grants ref EP/X035891/1 and EP/P022561/1.
For the purpose of open access, the author has applied a Creative Commons Attribution (CC BY) licence to any Author Accepted Manuscript version arising from this submission.  MK was supported by an eCSE grant from EPSRC. The authors have no conflicts to disclose.
\end{acknowledgments}

\end{document}

\maketitle	

\tableofcontents
\pagebreak

\section{Enthalpy vs pressure for DFT and MLIP.}
\begin{figure}[H]
\centering
\includegraphics[width=0.9\columnwidth]{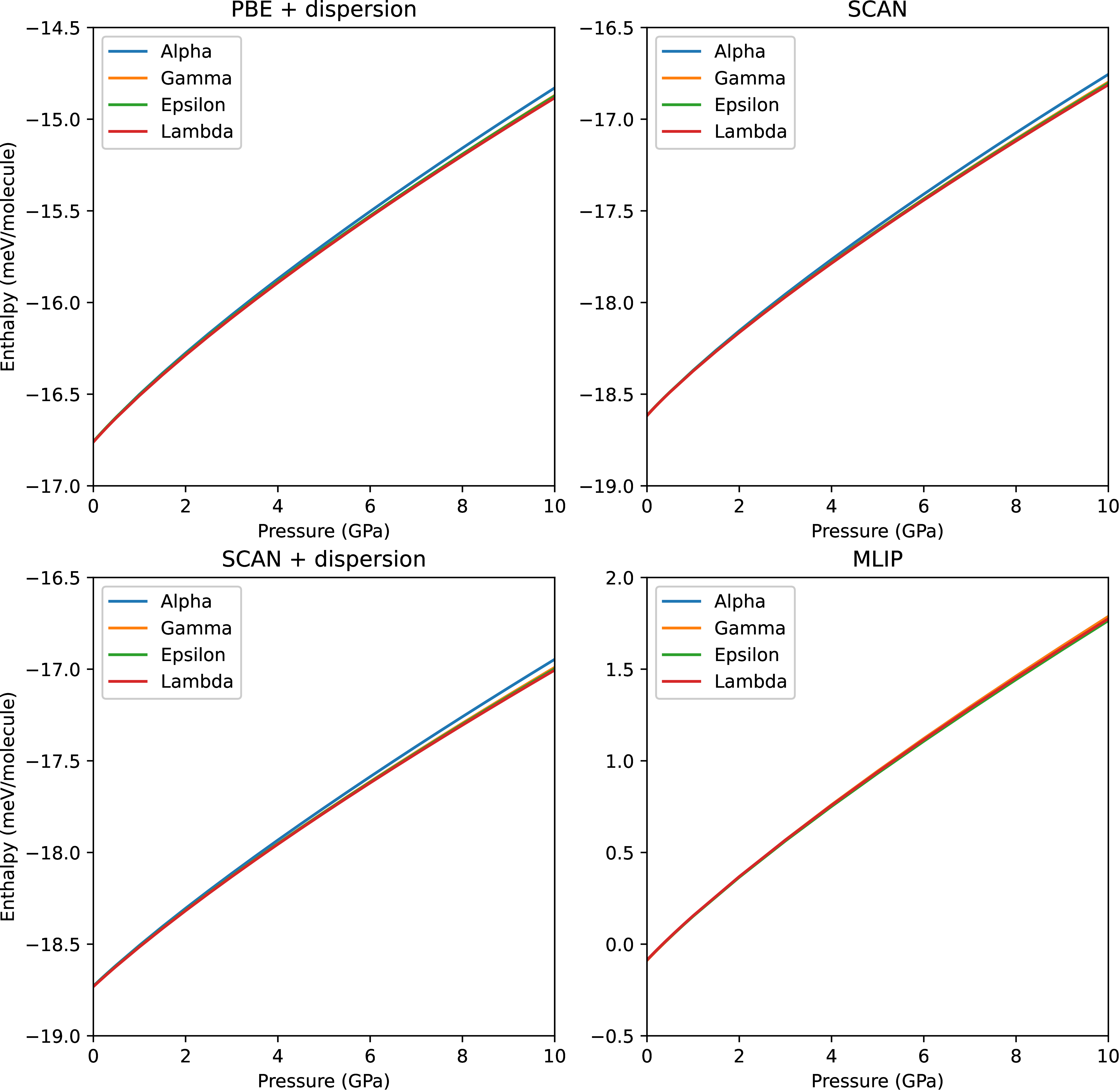}
\caption{Enthalpy vs pressure for  $\alpha$, $\gamma$, $\epsilon$ and $\lambda$ phases from static relaxation using the MLIP and from DFT with PBE+vdW, SCAN and SCAN+vdW functionals.}
\label{fig:GS}
\end{figure}

\section{Descriptors hyperparameters}
\begin{table}[H]
\centering
\begin{tabular}{ |c|c|c|c|c| } 
 \hline
 $r_s$ & 2.61066139097  &   0.10228258167   &  11.9012832034  &   1.56876280611 \\ 
 \hline
 $\eta$ & 1.02994852292  &   7.79076437642  &   0.04040235561  &   0.01340519269 \\ 
 \hline
\end{tabular}
\caption{Two-body blip function hyperparameters.}
\label{tab::hyperparams_2b}
\end{table}

\begin{table}[H]
\centering
\begin{tabular}{ |c|c|c|c|c| } 
 \hline
 $r_s$ & 0.00472139504  &   1.63499413497     & 0.00355931367   &  0.00361552429 \\ 
 \hline
 $\eta$ & 0.10854249660  &   0.78122061421   &  0.64121026653  &   0.31448274608 \\ 
 \hline
\end{tabular}
\caption{Many-body blip function hyperparameters.}
\label{tab::hyperparams_mb}
\end{table}

\section{Radial Distribution Function}
\begin{figure}[H]
\centering
\includegraphics[width=0.6\columnwidth]{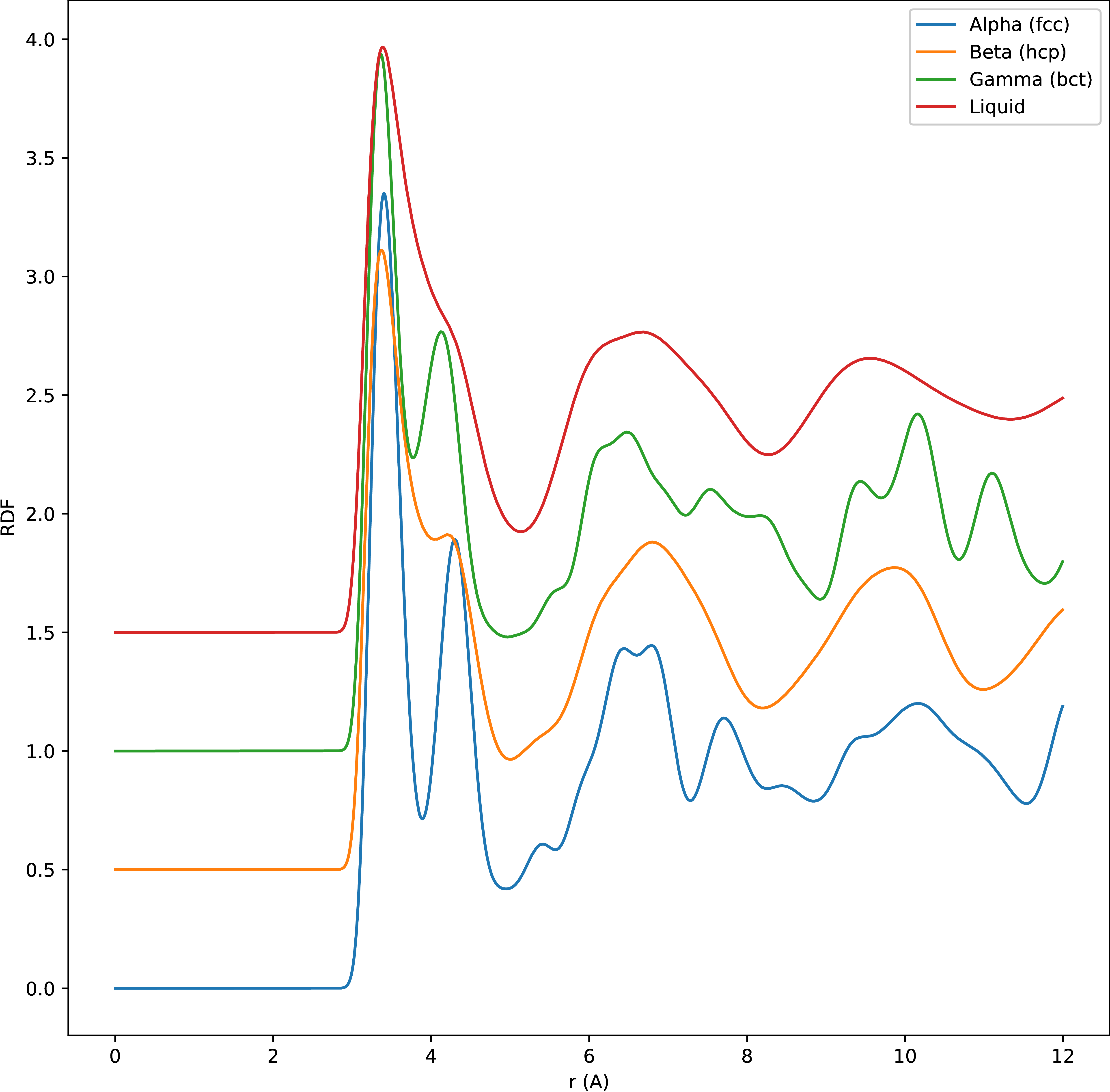}
\caption{Simulated (NPT) radial distribution functions for liquid and low pressure phases of molecular nitrogen.}
\end{figure}

\begin{figure}[H]
\centering
\includegraphics[width=0.48\columnwidth]{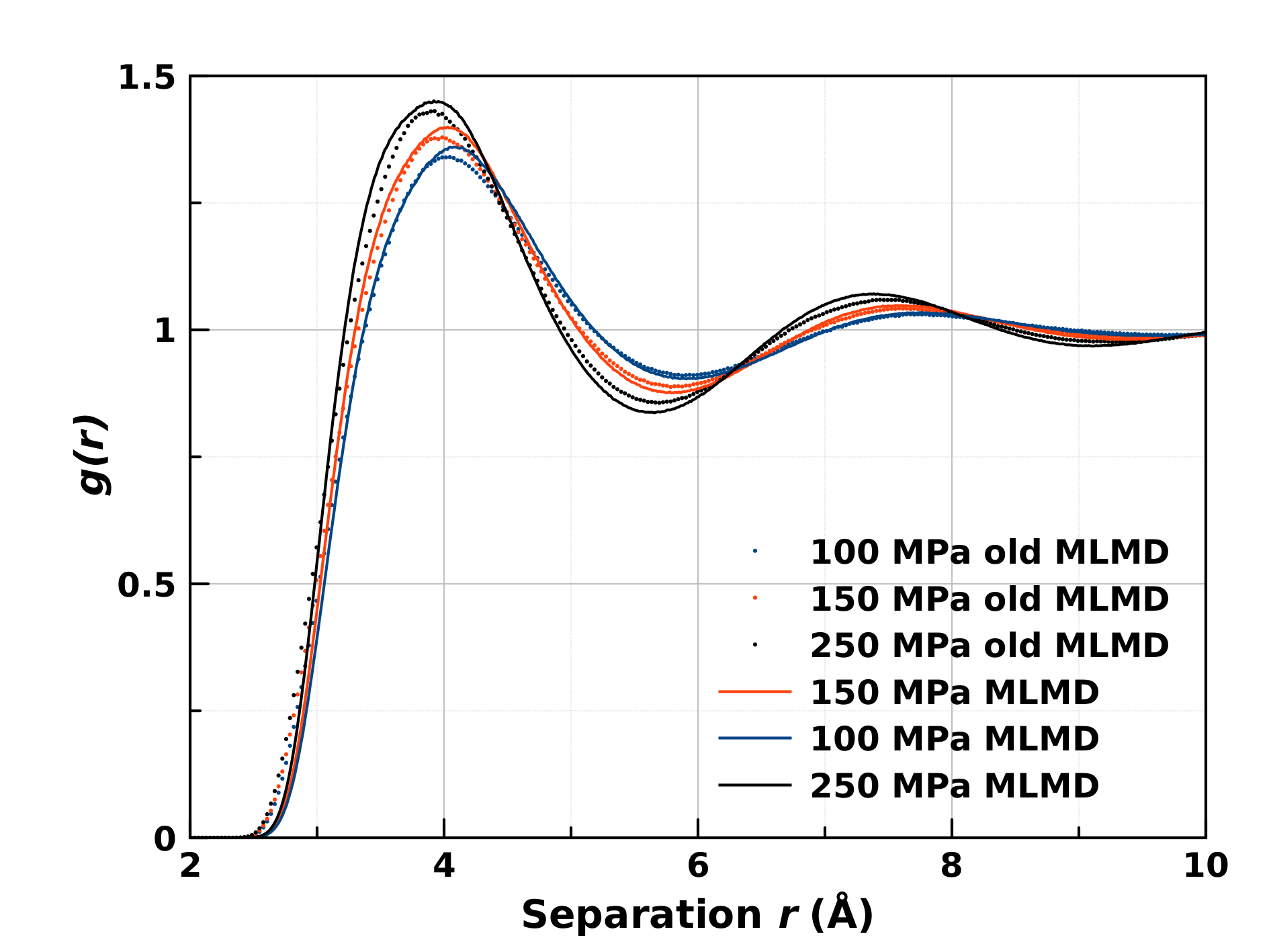}
\includegraphics[width=0.48\columnwidth]{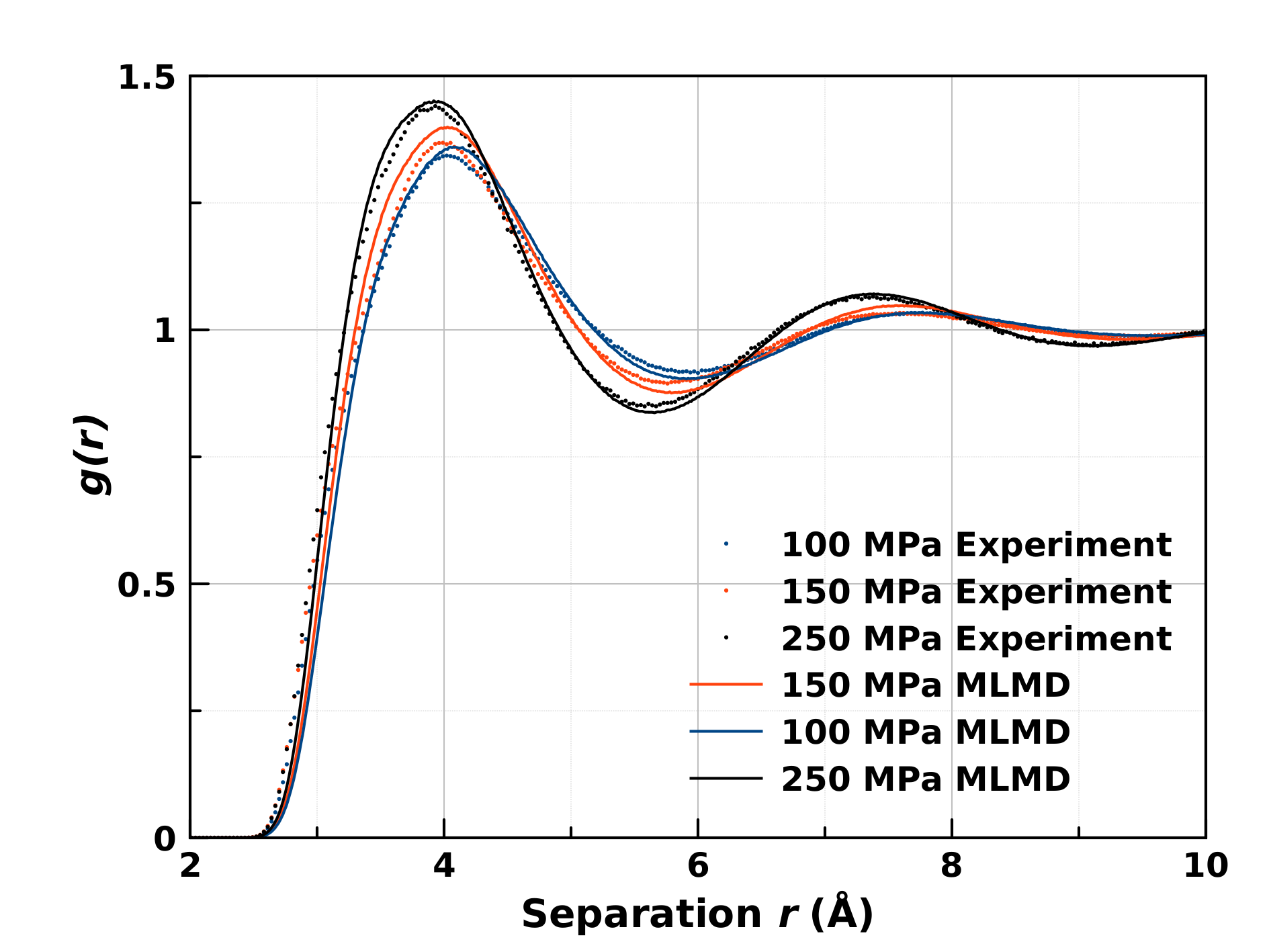}
\caption{(Left) Comparison of simulated RDFs between this model and model developed in \cite{Pruteanu2021}. (Right) RDFs generated with the model from this paper and experimentally obtained using neutron diffraction \cite{Pruteanu2021}. } 
\end{figure}

\section{Images of rotation}

We generate the orientational distribution of N2 molecules in the polar angles $\theta$, $\phi$ from the direction of the molecular axis in the MD runs.

At each snapshot in an MD trajectory, the molecular orientation in $\theta$ and $\phi$ was calculated. These values were then assigned to bins on the surface of a unit sphere. The bin centres were taken from an approximately uniform set of points on a sphere generated with a script adapted from the S$^2$ Sampling Toolbox in MATLAB \cite{MATLABpackage}.

\begin{figure}[H]
\label{fig:Blender}
\centering
\includegraphics[width=0.8\columnwidth]{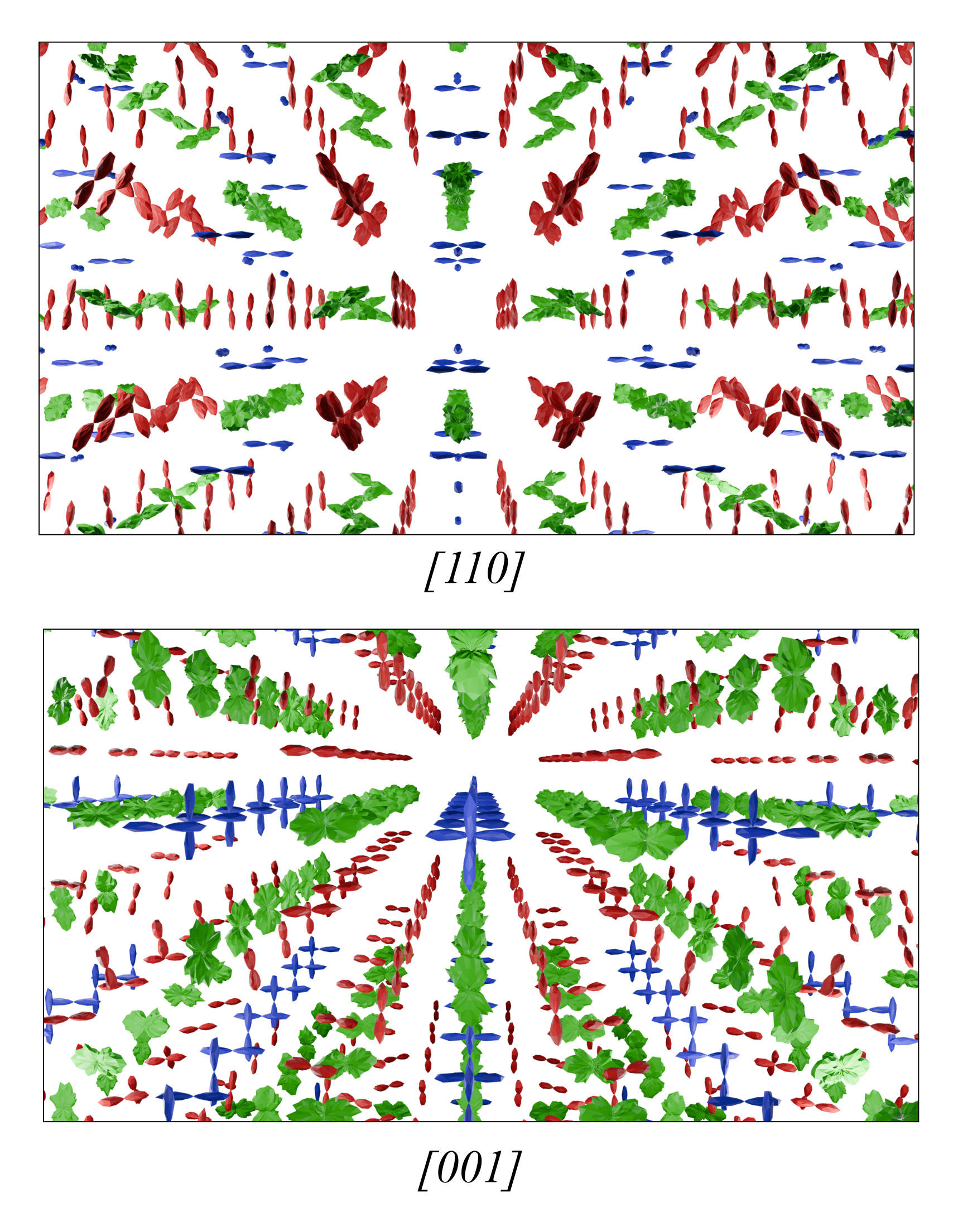}
\caption{Orientational PDFs for all molecules in the $\delta$* phase at 120 K and 10 GPa.}
\end{figure}

In Fig. \ref{fig:Blender}, the orientational PDFs are plotted as surfaces in $r,\theta,\phi$. The value of r corresponds to the probability density for a given orientation in $\theta,\phi$. The values of r are normalised for each molecule independently and scaled such that the maximum value of r is equal across all molecules.

The PDFs in Fig. \ref{fig:Blender} are positioned at the time averaged molecular centre of mass over the MD trajectory.

%of $\phi$ angles at the poles, interpreting these plots is not trivial. 
%These typically show a distribution about a well-defined direction  for molecules are not rotating. 
%In the case where the molecule points along the z-axis all values of $\phi$ are allowed, some on a square figure we see a band at $\theta= 0$ and $\pm \pi$
%For a 3D rotor, the histogram is flat with all solids angle having the same probability.  2D rotors are a little more complicated: rotation about the z-axis gives a horizontal band around $\theta=\pi/2$, rotating with an axis perpendicular to z gives a vertical band which splays out around  $\theta= 0$ and $\pm \pi$.  Other axes give diagonal bands.

\section{{N$_2$} Phases}
\begin{table*}[hb]
\begin{tabular}{cclllllll}
Pressure & Space group & \multicolumn{3}{c}{Lattice parameters} &
\multicolumn{4}{c}{Atomic coordinates} \\ (GPa) & &
\multicolumn{3}{c}{(\AA, $^{\circ}$)} &
\multicolumn{4}{c}{(fractional)} \\
\hline 0 & $P6_3/mmc$ & $a$=2.922 &$b$=4.17 & $c$=7.01 & N & 0.3333 & 0.6666 & 0.323 \\ & &
$\alpha$=90.00 & $\beta$=90 & $\gamma$=120.00 & & & & \\ & & & & &
& & & \\ 0 & Pa$\overline{3}$ & $a$=5.56 &
$b$=5.56 & $c$=5.56 & N & 0.0574	&0.0574	& 0.0574   \\ & &
$\alpha$=90.00 & $\beta$=90.00 & $\gamma$=90.00 & & & & \\ 
\\ 0 & $P2_1\bar3$ & $a$=5.56 &
$b$=5.56 & $c$=5.56 & N1 & 0.064	&0.064	& 0.064   \\ & &
$\alpha$=90.00 & $\beta$=90.00 & $\gamma$=90.00 & N2 & 
0.095& 0.095& 0.095  \\ & & & & &
& & & \\ 0 & $P2_1/c$ & $a$=2.922 &
$b$=2.891 & $c$=5.588 & N & 0.5678 & 0.3764 & 0.4534 \\ & &
$\alpha$=90.00 & $\beta$=132.54 & $\gamma$=90.00 & & & & \\ & & & & &
& & & \\
0 & $P4_2/mnm$ & $a$=4.08 &
$b$=4.08 & $c$=5.35 & N & 0.904 & 0.904 & 0 \\ & &
$\alpha$=90.00 & $\beta$=90.00 & $\gamma$=90.00 & & & & \\ & & & & &
& & & \\0 & $R\bar3c$ & $a$=9.86 &
$b$=9.86 & $c$=13.21 & N1 & 0.6666 & 0.3333 & 0 \\ & &
$\alpha$=90.00 & $\beta$=90.00 & $\gamma$=120.00 &N2 & 0.451 & 0.386 & 0.107  \\ & & & & &
& & & \\
\\ 40 & $Pbcn$ & $a$=2.895 & $b$=4.606 & $c$=5.221 & N &
0.1243 & 0.3576 & 0.8301 \\ & & $\alpha$=90.00 & $\beta$=90.00 &
$\gamma$=90.00 & & & & \\ & & & & & & & & \\\\ 40 & $P4_12_12$ &
$a$=2.907 & $b$=2.907 & $c$=8.228 & N & 0.4463 & 0.3019 & 0.7267 \\ &
& $\alpha$=90.00 & $\beta$=90.00 & $\gamma$=90.00 & & & & \\ & & & & &
& & & \\\\ 40 & $C2/c$ & $a$=4.112 & $b$=4.110 & $c$=9.795 & N1&
0.4041 & 0.6783 & 0.2732 \\ & & $\alpha$=90.00 & $\beta$=122.84 &
$\gamma$=90.00 & N2& 0.7789 & 0.3759 & 0.5233 \\ & & & & & & & & \\\\
56\cite{turnbull2018unusually} & $P2_1/c$ & $a$=9.899 & $b$=8.863 & $c$=8.726 & N1-N24& &
 &\\ & & $\alpha$=90.00 & $\beta$=91.64 & $\gamma$=90.00
& &  & &  \\ & & & & & & & &  \\\\
200 & $P\bar42_1m$ & $a$=3.009 & $b$=3.009 & $c$=4.029 & N1& 0.1645 &
0.3355 & 0.6815 \\ & & $\alpha$=90.00 & $\beta$=90.00 & $\gamma$=90.00
& N2& 0.5000 & 0.5000 & 0.8404 \\ & & & & & & & & \\\\ 200 &
$P2_12_12_1$ & $a$=2.569 & $b$=3.482 & $c$=4.019 & N1& 0.4909 & 0.4726
& 0.4150 \\ & & $\alpha$=90.00 & $\beta$=90.00 & $\gamma$=90.00 & N2&
0.7343 & 0.1478 & 0.3206 \\ & & & & & & & & \\\\
\end{tabular}
\caption{Structures from DFT, Materials Project and high pressure \cite{pickard2009high} }
\end{table*}

\clearpage
\bibliographystyle{plain}